\documentclass[preprint,preprintnumbers,amsmath,amssymb,floatfix,nofootinbib]{revtex4}
\usepackage{epsf}
\usepackage{graphicx}
\usepackage{subfigure}
\usepackage{array}
\usepackage{color}
\usepackage{appendix}
\usepackage{caption}
\newcommand{\be}{\begin{equation}}
\newcommand{\ee}{\end{equation}}

\title{Title}
\date{}
\begin{document}
\large
\date{\today}

\title{Lepton Flavor Specific Extended Higgs Model}

\author{B. L. Gon\c{c}alves}
\affiliation{Departamento de F\'{\i}sica and CFTP, Instituto Superior T\'ecnico, Universidade de Lisboa, Lisboa, Portugal}
\affiliation{Centro de F\'{\i}sica Te\'orica e Computacional, Faculdade de Ciências,
Universidade de Lisboa, Campo Grande, Edif\'{\i}cio C8, 1749-016 Lisboa, Portugal}
\author{Matthew Knauss}
\affiliation{High Energy Theory Group, William \& Mary, Williamsburg,
VA 23187, USA}
\author{Marc Sher}
\affiliation{High Energy Theory Group, William \& Mary, Williamsburg,
VA 23187, USA}


\begin{abstract}
In extended Higgs models, a discrete symmetry is needed in the quark sector to avoid tree-level flavor-changing neutral currents.   However, this is not necessarily the case in the lepton sector.    We consider a model in which one Higgs couples to quarks and three others couple to the electron, muon and tau, respectively.   This four-doublet model is presented with the full scalar potential and the gauge and Yukawa couplings.   The constraints from boundedness, perturbativity and oblique parameters are incorporated as well as constraints from meson-antimeson mixing, radiative $B$-decays and the diphoton Higgs decay rate. We also consider bounds from searches for heavy neutral and charged scalars at the LHC.   Since the Standard Model Higgs couplings match predictions very well, we focus on the alignment limit of the model.   It is shown that for a wide range of parameters, the lightest additional scalar, pseudoscalar and charged scalar can have substantial decays into electrons and muons (in contrast to the usual leptonic decays into taus).   An interesting signature in the neutral sector would be the production, through vector boson fusion, of a pair of scalars, each of which decays into an electron or muon pair.
\end{abstract}

\maketitle

\section{Introduction}

The Higgs boson was initially discovered \cite{Aad:2012tfa,Chatrchyan:2012xdj} through its decay 
into gauge bosons.    Since then, the coupling of the Higgs to third generation fermions has also 
been determined with increasing accuracy \cite{Aad:2015vsa, Chatrchyan:2014nva, Aaboud:2018zhk, 
CMS:2018nsn, CMS:2018uxb}.    

 However, while there is evidence \cite{CMS:2020xwi} of the Higgs decay into muons, there remain large uncertainties and the discovery has not yet
 been made.    This leads one to ask if there are viable models in which the muon and tau couple to different Higgs bosons.
 It is often claimed that models in which fermions of a given charge couple to different Higgs bosons contain tree-level flavor changing neutral currents (FCNC).   However, the seminal papers of Glashow and Weinberg~\cite{Glashow:1976nt} and of Paschos~\cite{Paschos:1976ay} explicitly referred to the quark sector.    As we will see, FCNC can be avoided in the lepton sector even if different leptons couple to different Higgs bosons.
 
 The first such model,  called the muon-specific Two
Higgs Doublet (2HDM) model, was developed by Abe, Sato and Yagyu \cite{Abe:2017jqo} (ASY).   They use a
$Z_4$ symmetry, under which the muon and tau have different quantum numbers, and break this softly.     The
model has no tree-level FCNC and the Yukawa couplings for the muon and tau are no longer simply
proportional to their masses with the proportionality coefficient being the same for all flavors: 
rather, the ASY model can substantially enhance or suppress the muon interactions of scalars relative 
to those with tau leptons. The purpose
of their model was to attempt an explanation of the muon g-2 anomaly, and for the parameters they considered,
the dimuon coupling of the 125 GeV Higgs is not suppressed.   Their model can address the g-2 anomaly, but only for a very narrow region of parameter-space.
A more detailed analysis was carried out in Ref. \cite{Ferreira:2020ukv} where the phenomenology of the model was studied.

The ASY muon specific 2HDM used a $Z_4$ discrete symmetry in which the left-handed muon doublet and right-handed singlet have charge $i$ and $\Phi_1$ has charge -1.  All other fields have charge +1.   This then has $\Phi_1$ coupling to muons and $\Phi_2$ coupling to all other fermions.     Ivanov and Nishi have pointed out~\cite{Ivanov:2013bka} that the actual symmetry group of the model is a softly broken $Z_2$ in which  $\Phi_1$ and $\mu_R$ are negative and with a $U(1)$ corresponding
to muon number.   This does not affect the ASY Lagrangian.    In this model, the mass matrix of the charged leptons breaks into a $2\times 2$ submatrix, corresponding to $e-\tau$ and a $1\times 1$ corresponding to the muon.   One might be concerned about how the PMNS matrix is generated if the muon and muon neutrino mass matrices decouple.   However, even if the charged lepton and neutrino mass matrices are diagonal, one will still obtain a PMNS matrix using the see-saw (type 1) mechanism.   The light neutrino mass matrix is then $m_{ij} = (M_D)_{ik} (M_N)^{-1}_{kl} (M_D)_{lj}$ where $M_D$ is the diagonal Dirac neutrino mass matrix and $M_N$ is the superheavy Majorana right-handed neutrino mass matrix.  The latter is arbitrary and so the light neutrino mass matrix is not diagonal, leading to a non-trivial PMNS matrix.   Note that this will not work in the quark sector.

In this paper, we take the ASY model one step further and suppose that {\it each }of the charged leptons couples to a different Higgs doublet, which we will label as  $\Phi_e, \Phi_\mu$ and $\Phi_\tau$.    This can be achieved with a $(Z_4)_e \times (Z_4)_\mu \times (Z_4)_\tau$ symmetry in which $L_\ell$ and $\ell_R$ have quantum number under $(Z_4)_\ell$ of $i$ and the $\Phi_\ell$ has quantum number $-1$.     Equivalently, one can replace the $Z_4$ with $Z_2\times U(1)$ as discussed above - the Lagrangian in either case is identical.    To achieve a non-trivial PMNS matrix, the symmetry must be softly broken in the superheavy Majorana neutrino mass matrix.     The simplest implementation of this model would be a 4HDM in which the fourth Higgs $\Phi_q$ couples to the quarks.    This is similar to the lepton-specific model.     Certainly one could have one of $\Phi_\ell$ be the same as $\Phi_q$, leading to a 3HDM.    However, if the $\Phi_q$ is $\Phi_\tau$, then the resulting model is very similar to the muon-specific model - the only difference  being the very small interaction of the Higgs with the electron.    For simplicity, we assume they are separate.   One could also adopt a type-II structure, with 5HDM, but that brings in additional complications and the type-II parameter space is much narrower than the type-I.    So, we will focus on the 4HDM with $\Phi_q, \Phi_e,\Phi_\mu$ and $\Phi_\tau$.

Although there are hundreds of papers that study models with three Higgs doublets, very few look at models with four.    A recent paper with 4HDM in which each Higgs couples to sets of fermions with similar masses has been proposed \cite{Rodejohann:2019izm} and a special ansatz, ``singular alignment", is needed to suppress FCNC. A supersymmetric model \cite{Arroyo-Urena:2019lzv} had one doublet each coupling to up-quarks, down-quarks and leptons, with the fourth needed for anomaly cancellation.   A similar non-supersymmetric model was proposed \cite{Cree:2011uy}(with the fourth Higgs needed to relax some tight constraints).    An early discussion that mentions 4HDMs  \cite{Ivanov:2011ae} studied Abelian symmetries in multidoublet models.  There are also many studies of symmetries and vacuum states of $N$ doublet models.   An extremely extensive 2017 review of Ivanov \cite{Ivanov:2017dad}, with over 500 references, studied numerous extended scalar sectors (including two doublet models, $N$ doublet models, singlet and triplet extensions).   Most relevant papers before that time are referred to in this review.  A more recent paper \cite{Faro:2020qyp} looked at the interesting issue of non-decoupling in multiscalar models.  Related work \cite{deMedeirosVarzielas:2021zqs} dealt with large discrete symmetry groups in $N$ doublet models.    Additionally, the ``Private Higgs" model of Porto and Zee \cite{Porto:2007ed,Porto:2008hb} had one Higgs doublet for every fermion.   In contrast to the model we propose, their model had numerous discrete symmetries and included several ``darkon" scalars.

We see that there are many models with more than two Higgs doublets in the literature.   All of these treat the three charged leptons identically, except for the ``Private Higgs" model which treated quarks and leptons in the same manner.   Yet the lepton sector is one of the most mysterious, given the large mixing angles and small masses in the neutrino sector.   In this paper, we are treating the lepton and quark sectors differently, but not treating the charged leptons identically, coupling each lepton to a separate Higgs.

In section II, the model is presented, including the full scalar potential and the gauge and Yukawa couplings.   In section III, we discuss the constraints on the potential from boundedness and constraints from oblique parameters.  In section IV, two benchmark models are presented.  In the first model, the potential is divided into two $2 \times 2$ subsections and in the second, the full $4 \times 4$ model is discussed in the experimentally indicated alignment limit.   Section V contains our results and conclusions.

\section{The Model}

\subsection{Scalar sector}

The potential can be written as a sum of quadratic and quartic terms:  $V = V_2 + V_4$.   We allow for soft breaking of the discrete symmetry in the quadratic terms:
\be
\begin{aligned}
V_2 = & \ m_{qq}^2 \Phi_q^\dagger\Phi_q  + m_{ee}^2 \Phi_e^\dagger\Phi_e + m_{\mu\mu}^2 \Phi_\mu^\dagger\Phi_\mu + m_{\tau\tau}^2 \Phi_\tau^\dagger\Phi_\tau\\ +
& \ [m_{qe}^2(\Phi_q^\dagger \Phi_e) + m_{q\mu}^2(\Phi_q^\dagger \Phi_\mu) + m_{q\tau}^2(\Phi_q^\dagger \Phi_\tau)\\ + & \ m_{e\mu}^2(\Phi_e^\dagger \Phi_\mu) + m_{e\tau}^2(\Phi_e^\dagger \Phi_\tau) + m_{\mu\tau}^2(\Phi_\mu^\dagger \Phi_\tau)] + {\rm h.c.} \end{aligned}
\ee
and
\be \begin{aligned}
V_4 =& \ \lambda_1^q(\Phi_q^\dagger\Phi_q)^2 +  \lambda_1^e(\Phi_e^\dagger\Phi_e)^2 +  \lambda_1^\mu(\Phi_\mu^\dagger\Phi_\mu)^2 +  \lambda_1^\tau(\Phi_\tau^\dagger\Phi_\tau)^2\\ + &\ 
 \lambda_3^{qe}(\Phi_q^\dagger\Phi_q)(\Phi_e^\dagger\Phi_e) +  \lambda_3^{q\mu}(\Phi_q^\dagger\Phi_q)(\Phi_\mu^\dagger\Phi_\mu) +  \lambda_3^{q\tau}(\Phi_q^\dagger\Phi_q)(\Phi_\tau^\dagger\Phi_\tau) \\ +&   \lambda_3^{e\mu}(\Phi_e^\dagger\Phi_e)(\Phi_\mu^\dagger\Phi_\mu) +  \lambda_3^{e\tau}(\Phi_e^\dagger\Phi_e)(\Phi_\tau^\dagger\Phi_\tau) +  \lambda_{3}^{\mu\tau}(\Phi_\mu^\dagger\Phi_\mu)(\Phi_\tau^\dagger\Phi_\tau)
 \\ + &\ \lambda_{4}^{qe}(\Phi_q^\dagger\Phi_e)(\Phi_e^\dagger\Phi_q) +  \lambda_{4}^{q\mu}(\Phi_q^\dagger\Phi_\mu)(\Phi_\mu^\dagger\Phi_q) +  \lambda_{4}^{q\tau}(\Phi_q^\dagger\Phi_\tau)(\Phi_\tau^\dagger\Phi_q) \\ +&\   \lambda_{4}^{e\mu}(\Phi_e^\dagger\Phi_\mu)(\Phi_\mu^\dagger\Phi_e) +  \lambda_{4}^{e\tau}(\Phi_e^\dagger\Phi_\tau)(\Phi_\tau^\dagger\Phi_e) +  \lambda_{4}^{\mu\tau}(\Phi_\mu^\dagger\Phi_\tau)(\Phi_\tau^\dagger\Phi_\mu)\\ +&\  
 \frac{1}{2}\big[ \lambda_{5}^{qe}(\Phi_q^\dagger\Phi_e)^2+  \lambda_5^{q\mu}(\Phi_q^\dagger\Phi_\mu)^2 +  \lambda_{5}^{q\tau}(\Phi_q^\dagger\Phi_\tau)^2 \\ +&\   \lambda_{5}^{e\mu}(\Phi_e^\dagger\Phi_\mu)^2+  \lambda_{5}^{e\tau}(\Phi_e^\dagger\Phi_\tau)^2 +  \lambda_{5}^{\mu\tau}(\Phi_\mu^\dagger\Phi_\tau)^2+ {\rm h.c.}\big]
 \end{aligned}
 \ee
Here, we have labeled the quartic couplings to be similar to the standard 2HDM potential.

 If the $m_{{ij}}^2$ and $\lambda_5^{ij}$ have imaginary components, one would have CP violation in the scalar sector.   There are six CP violating parameters from the $m_{ij}^2$ and another six from the $\lambda_5^{ij}$ parameters.  Two can be eliminated through rescaling, but that would leave ten additional parameters.  A detailed analysis of CP violation in the 3HDM \cite{Akeroyd:2021fpf} considered the effects on the neutron and electron electric dipole moments as well as CP violating effects in B decays.  They also discuss mixing of scalars and pseudoscalars which would complicate the analysis.    In our model, one would expect very similar effects.    For simplicity, we will assume that these parameters are real and refer the reader to Ref. \cite{Akeroyd:2021fpf} for details.

We can write the Higgs doublets as
\be
\Phi_i = \begin{pmatrix} \phi^+_i \\ (v_i + \phi_i + i\chi_i)/\sqrt{2} \end{pmatrix}, (i=q,e,\mu,\tau)
\ee
where the $v_i/\sqrt{2}$ are the vacuum values of the neutral components.   To discuss diagonalizing mass matrices and the various angles involved, we follow the procedure of Boto, Rom\~ao and Silva~\cite{Boto:2021qgu} closely.

Without loss of generality, we can define the angles that rotate the fields into the Higgs basis in which only one scalar field gets a vev by
\be\begin{aligned}
v_q &= v\cos\beta_2\cos\beta_3\cos\beta_4\\ v_e &= v\sin\beta_2\cos\beta_3\cos\beta_4\\ v_\mu &= v\sin\beta_3\cos\beta_4\\  v_\tau& = v\sin\beta_4
\end{aligned}\ee
giving
\be
\begin{pmatrix} h_0\\ H_1\\ H_2 \\H_3 \end{pmatrix} = {\cal O}_\beta \begin{pmatrix} \phi_q\\ \phi_e \\ \phi_\mu \\ \phi_\tau\end{pmatrix} \ee
where  
\be
\mathcal{O}_\beta = \begin{pmatrix} 
c_{\beta_2}c_{\beta_3}c_{\beta_4} & s_{\beta_2}c_{\beta_3}c_{\beta_4} & s_{\beta_3}c_{\beta_4} & s_{\beta_4} \\ -s_{\beta_2} & c_{\beta_2} & 0 & 0 \\ -c_{\beta_2}c_{\beta_3} & -s_{\beta_2}s_{\beta_3} & c_{\beta_3} & 0 \\ -c_{\beta_2}c_{\beta_3}s_{\beta_4} & -s_{\beta_2}c_{\beta_3}s_{\beta_4} & -s_{\beta_3}s_{\beta_4} & c_{\beta_4}
\end{pmatrix}
\ee
Here, $h_0$ is the field that gets the entire vev, $v$, and $c_\theta$ ($s_\theta$) are $\cos\theta$ ($\sin\theta$).  

From this basis, we can now diagonalize the mass matrices of the various scalars.   In the neutral scalar sector, the physical neutral Higgs masses are given by
\be
\begin{pmatrix} h_1\\ h_2\\ h_3 \\h_4 \end{pmatrix} = {\cal O}_\alpha \begin{pmatrix} \phi_q\\ \phi_e \\ \phi_\mu \\ \phi_\tau\end{pmatrix} \ee
where $h_1$ is the 125 GeV Higgs particle.    For ${\cal O}_\alpha$, we use 
\be
{\cal O}_\alpha = {\bf R}_{34}{\bf R}_{24}{\bf R}_{23}{\bf R}_{14}{\bf R}_{13}{\bf R}_{12}\ee
Here, for example, ${\bf R}_{24}$ is given by
\be
{\bf R}_{24} = \begin{pmatrix} 1&0&0&0\\ 0&c_{\alpha_{24}} & 0 & s_{\alpha_{24}}\\ 0&0&1&0\\ 0&-s_{\alpha_{24}}&0&c_{\alpha_{24}} \end{pmatrix}
\ee 
and the other R matrices follow.   We see that there are six rotation angles.

In the pseudoscalar sector, one has
\be
\begin{pmatrix} 
G^0\\ A_1 \\ A_2 \\ A_3 \end{pmatrix} = {\cal O}_\gamma {\cal O}_\beta \begin{pmatrix} \chi_q \\ \chi_e \\ \chi_\mu \\ \chi_\tau \end{pmatrix}
\ee
where ${\cal O}_\gamma = {\bf P}_{34} {\bf P}_{24} {\bf P}_{23}$ and, as before, for example
\be
{\bf P}_{24} = \begin{pmatrix} 1&0&0&0 \\ 0& c_{\gamma_{24}} &0 & s_{\gamma_{24}}\\ 0&0&1&0\\ 0 & -s_{\gamma_{24}} & 0 & c_{\gamma_{24}} \end{pmatrix}
\ee
Note that there are only three matrices here, since the Goldstone boson direction is fixed.

Finally, in the charged sector
\be
\begin{pmatrix} 
G^+\\ H_1^+ \\ H^+_2 \\ H^+_3 \end{pmatrix} = {\cal O}_\delta {\cal O}_\beta \begin{pmatrix} \phi^+_q \\ \phi^+_e \\ \phi^+_\mu \\ \phi^+_\tau \end{pmatrix}
\ee
where ${\cal O}_\delta = {\bf Q}_{34} {\bf Q}_{24} {\bf Q}_{23}$ and, as before, for example
\be
{\bf Q}_{24} = \begin{pmatrix} 1&0&0&0 \\ 0& c_{\delta_{24}} &0 & s_{\delta_{24}}\\ 0&0&1&0\\ 0 & -s_{\delta_{24}} & 0 & c_{\delta_{24}} \end{pmatrix}
\ee

\subsection{Gauge and Yukawa couplings}

\subsubsection{Gauge couplings}
The scalar kinetic Lagrangian, $\mathcal{L}_\text{k}$, defined as
\be
\begin{split}
\mathcal{L}_\text{k}=\sum_{i=1}^{4} |D_{\mu}\Phi_{i}|^{2} 
\end{split}
\ee
with the usual expression for the covariant derivative $D_\mu$, contains the terms relevant to obtain the trilinear couplings of the scalars and gauge bosons. The couplings $ZZh_i$ and $W^{\pm}W^{\mp}h_i$ are written in the form 
\be
\begin{split}
\left(\sum_{i=1}^{4} C_i h_i\right)\left(\frac{g}{2 c_W}m_Z Z_\mu Z^\mu + g m_W W^{-}_\mu {W^{+}}^\mu  \right) \, .
\end{split}
\label{eq:gauge-couplings}
\ee
The $C_i$ factors are included in Appendix~\ref{label:appendixA}. It is possible to check that, when the set of conditions $\alpha_{1j}= \beta_j$ is verified (for $j=2,3,4$), one gets $C_1=1$ together with $C_k=0$, for $k \neq 1$, which defines the alignment limit in this model.


\subsubsection{Yukawa couplings}

Following the notation of Branco, et al.~\cite{Branco:2011iw}, the couplings of the scalar and pseudoscalar Higgs are defined through
\be
\begin{split}
\mathcal{L}_{Y}^S =& - \sum_{f\in \left\{q,e,\mu,\tau\right\}} \frac{m_f}{v} \left( \xi_{h_1}^{f} \bar{f}f h_1 + \xi_{h_2}^{f} \bar{f}f h_2 + \xi_{h_3}^{f} \bar{f}f h_3 + \xi_{h_4}^{f} \bar{f}f h_4 \right) \\
\mathcal{L}_{Y}^P =& -\sum_{f\in \left\{q,e,\mu,\tau\right\}} \left(-i\frac{m_f}{v}\right) \left( \xi_{A_1}^{f} \bar{f}\gamma_5 f A_1 + \xi_{A_2}^{f} \bar{f}\gamma_5 f A_2 + \xi_{A_3}^{f} \bar{f}\gamma_5 f A_3 \right)
\end{split}
\label{eq:yukawas-scalar-pseudoscalar-def1}
\ee
where $\xi_{h_j}^f$ and $\xi_{A_j}^f$ are given by 
\be
\begin{split}
\xi_{h_j}^q = \frac{{\cal{O}_\alpha}_{j,1}}{\hat{v}_1} \: ,\: \xi_{h_j}^e = \frac{{\cal{O}_\alpha}_{j,2}}{\hat{v}_2} \: ,& \: \xi_{h_j}^\mu = \frac{{\cal{O}_\alpha}_{j,3}}{\hat{v}_3} \: , \: \xi_{h_j}^\tau = \frac{{\cal{O}_\alpha}_{j,4}}{\hat{v}_4}  \\
\xi_{A_j}^q = \frac{\left({\cal{O}_\gamma\cal{O}_\beta}\right)_{j,1}}{\hat{v}_1} \: ,\: \xi_{A_j}^e = \frac{\left({\cal{O}_\gamma\cal{O}_\beta}\right)_{j,2}}{\hat{v}_2} \: ,& \: \xi_{A_j}^\mu = \frac{\left({\cal{O}_\gamma\cal{O}_\beta}\right)_{j,3}}{\hat{v}_3} \: , \: \xi_{A_j}^\tau = \frac{\left({\cal{O}_\gamma\cal{O}_\beta}\right)_{j,4}}{\hat{v}_4}
\end{split}
\label{eq:yukawas-scalar-pseudoscalar-def2}
\ee
using $\hat{v}_i \equiv v_i / v$. 
Similarly, the couplings of the charged Higgs are defined through
\be
\begin{split}
    \mathcal{L}_{Y}^C =-\sum_{j}\Biggl[ \Biggr. &\left.\sum_{u,d}\frac{\sqrt{2}V_{ud}}{v}\bar{u}\left( m_u \xi_{H_j^+}^{qL} \text{P}_L + m_d \xi_{H_j^+}^{qR} \text{P}_R  \right)d H_j^+ \right. \\
    &\left. + \sum_{l} \frac{\sqrt{2}m_l}{v} \xi_{H_j^+}^{lL}\bar{\nu}_L l_R H_j^+ \right] + \text{h.c.} 
\end{split}
\label{eq:yukawas-charged-def1}
\ee
where $\xi_{H_j^+}^f$ are given by 
\be
\xi_{H_j^+}^{qLR} = \frac{\left({\cal O}_\delta {\cal O}_\beta\right)_{j,1}}{\hat{v}_1} \: ,\: \xi_{H_j^+}^{eL} = \frac{\left({\cal O}_\delta {\cal O}_\beta\right)_{j,2}}{\hat{v}_2} \: , \: \xi_{H_j^+}^{\mu L} = \frac{\left({\cal O}_\delta {\cal O}_\beta\right)_{j,3}}{\hat{v}_3} \: , \: \xi_{H_j^+}^{\tau L} = \frac{\left({\cal O}_\delta {\cal O}_\beta\right)_{j,4}}{\hat{v}_4} 
\label{eq:yukawas-charged-def2}
\ee

A table of general Yukawa couplings are included in Appendix~\ref{label:appendixB}. 

\section{Theoretical constraints on the scalar potential}

\subsection{Bounded from below constraints}

In extensions of the scalar sector, one needs to choose quartic parameters such that the potential is bounded from below (BFB)\footnote{ We require that the potential be bounded at scales where the quartic terms dominate.  The case in which the potential turns over at very high scales due to renormalization group running will not be considered.  In fact, the Standard Model itself would not satisfy that latter condition}.      While this is straightforward in the 2HDM, it can be quite complicated in models with more than two doublets.    An added complication in models with doublets is that there can be an instability in the charged scalar direction even if there is stability in the neutral scalar direction (see Ref. \cite{Faro:2019vcd} for an example).
A recent discussion of these conditions for a three-doublet model can be found in the work of Boto, Rom\~ao and Silva~\cite{Boto:2022uwv}.    They showed that while necessary and sufficient conditions are known for the neutral direction, only sufficient conditions are known for stability in the charged direction, and they discuss a general strategy.    We will first discuss the neutral directions.

Looking at the neutral direction, the 2HDM potential can be written as  $V_4 = a_{11} H_1^4 + a_{22} H_2^4 + a_{12} H_1^2 H_2^2$, where the matrix is symmetric.    The conditions for copositivity (where the potential is positive for all values of $H_1^2$ and $H_2^2$) are given by $a_{11} \geq 0, a_{22} \geq 0, a_{12} + \sqrt{a_{11}a_{22}} \geq 0$.    As shown in Refs.~\cite{Klimenko:1984qx,Kannike:2012pe}, for the neutral sector of the 3HDM, the conditions are
\begin{align}
a_{11} \geq 0,\quad a_{22} &\geq 0, \quad a_{33} \geq 0\\
a_{12} + \sqrt{a_{11}a_{22}} &\geq 0 \\
a_{13} + \sqrt{a_{11}a_{33}} &\geq 0 \\
a_{23} + \sqrt{a_{22}a_{33}} &\geq 0 \\
\sqrt{a_{11}a_{22}a_{33}} + a_{12}\sqrt{a_{33}} &+ a_{13}\sqrt{a_{22}} + a_{23}\sqrt{a_{11}} \geq 0\\
\det{A} & \geq 0
\end{align}
where $A$ is the matrix with entries $a_{ij}$.
Clearly, the first line is needed for stability along the axes, the next three lines are needed for stability in the three planes, and the last two lines ensure stability for all directions.
For the 4HDM that we consider, the corresponding conditions must be satisfied for every three dimensional subspace.   The remaining conditions are extremely complicated, but are given in full in Ref.~\cite{Klimenko:1984qx}. We have incorporated the conditions in that paper to ensure stability in the neutral directions.

As shown by Boto, Rom\~ao and Silva~\cite{Boto:2022uwv}, even in the 3HDM there are no straightforward necessary and sufficient conditions for stability in the charged directions.   In the 2HDM, with a quartic potential
\begin{equation}
V_4 = \lambda_1 (\Phi_1^\dagger \Phi_1)^2 +\lambda_2 (\Phi_2^\dagger \Phi_2)^2 + \lambda_3 (\Phi_1^\dagger \Phi_1)(\Phi_2^\dagger\Phi_2) + \lambda_4 |\Phi_1^\dagger \Phi_2|^2 + \frac{1}{2}\lambda_5[(\Phi_1^\dagger \Phi_2)^2 + (\Phi_2^\dagger\Phi_1)^2]
\end{equation}
the condition for stability is \cite{Ivanov:2015mwl,Song:2022gsz} $\lambda_3 + \lambda_4 -|\lambda_5| \geq -2\sqrt{\lambda_1\lambda_2}$.    Rather than attempt a detailed numerical study of stability in the 4HDM case, we will require that this condition be satisfied for all $2\times 2$ subspaces of the 4HDM.   This requirement is, of course, necessary but may not be sufficient.

\subsection{Oblique Parameters}
To discuss the S, T, U oblique parameters, we follow the methods and results in Grimus, et al \cite{Grimus:2008nb}. To do this, we can write the matrices $\tilde{\mathcal{U}}$ and $\tilde{\mathcal{V}}$ from Grimus, et al \cite{Grimus:2008nb} using our notation in the previous section. $\tilde{\mathcal{V}}$ is defined through
\be \label{vTildeDef}
\begin{pmatrix} \phi_1 \, +\, i\chi_1 \\ \phi_2 \,+\, i\chi_2 \\ \phi_3 \,+\, i\chi_3 \\ \phi_4 \,+\, i\chi_4 \end{pmatrix} = \tilde{\mathcal{V}} \begin{pmatrix} h_1 & h_2 & h_3 & h_4 & G^0 & A_1 & A_2 & A_3 \end{pmatrix}^T
\ee
where 
\be
\tilde{\mathcal{V}} \equiv \begin{pmatrix}
\mathcal{O}_\alpha^{-1} \\
i\left(\mathcal{O}_\gamma \mathcal{O}_\beta\right)^{-1}
\end{pmatrix}
\ee
Notice in Eq.~(\ref{vTildeDef}), our notation slightly differs from Grimus et al \cite{Grimus:2008nb} by keeping the Goldstone boson with the pseudoscalar mass eigenstates.

$\tilde{\mathcal{U}}$ is defined as
\be
\begin{pmatrix} \phi_1^+ \\ \phi_2^+ \\ \phi_3^+ \\ \phi_4^+ \end{pmatrix} = \tilde{\mathcal{U}} \begin{pmatrix} G^+ \\ H_1^+ \\ H_2^+ \\H_3^+ \end{pmatrix}
\ee
where 
\be
\tilde{\mathcal{U}} \equiv \begin{pmatrix}{\cal O}_\delta {\cal O}_\beta\end{pmatrix}
\ee

We take the values of S, T from \cite{Workman:2022ynf} with 
\be
\begin{aligned} \
S &= &-0.02 \pm 0.10 \\ T &= &0.03 \pm 0.12
\label{eq:oblique-values}
\end{aligned}
\ee

We will not include the detailed calculation of the unitarity and perturbativity bounds, due to the large number of scalar couplings.    Rather, we will simply require that all of the quartic scalar couplings be less than $4\pi$.   

\section{Benchmark models}

As is clear from examining the scalar potential and the Appendices, the model contains a large number of free parameters.    To focus on the most important aspects of the model, we will consider two benchmark models.    In the first, we will assume that the $(q\tau)$ sector of the Higgs potential decouples from the $(\mu e)$ sector.   In that case, the $4\times 4$ scalar mass matrices decouple into two $2\times 2$ matrices which can be trivially diagonalized analytically.    In the second benchmark model, we will take the alignment limit.   In the conventional 2HDMs, this is equivalent to $\cos(\alpha-\beta)=0$, with $\tan\beta \equiv v_2/v_1$ and $\alpha$ diagonalizes the scalar mass matrix.   This limit is often chosen since it means that the couplings of the $125$ GeV Higgs boson are identical to that in the Standard Model (which seems to be preferred by LHC data).     In this case, it is easy to see from Appendices~\ref{label:appendixA} and~\ref{label:appendixB} that the alignment limit corresponds to $\alpha_{1j}= \beta_j$, as  previously stated. Since the coupling of the $125$ GeV Higgs is the same as the Standard Model, there is no need to study Higgs production and tree-level decays in this case.

\subsection{The Model without $(q\tau)$-$(\mu e)$ mixing}

In this model, the absence of $(q\tau)$-$(\mu e)$ mixing means that the matrix that diagonalizes the scalar mass matrix, $\cal{O}_\alpha$, is broken into two $2\times 2$ matrices.  The upper $2\times 2$ matrix looks very similar to the lepton-specific 2HDM.   The only difference involves the coupling to the muon, which is not well-measured.   However, in this case, unlike the lepton-specific model, the value of $v_q^2 + v_\tau^2$ is not $v^2=(246\ \rm{GeV})^2$ but will be smaller.   As a result, all Yukawa couplings will be increased.   This will affect the decays of the $125$ GeV Higgs boson as well as the production.

\begin{figure}
\includegraphics[width=9.05cm]{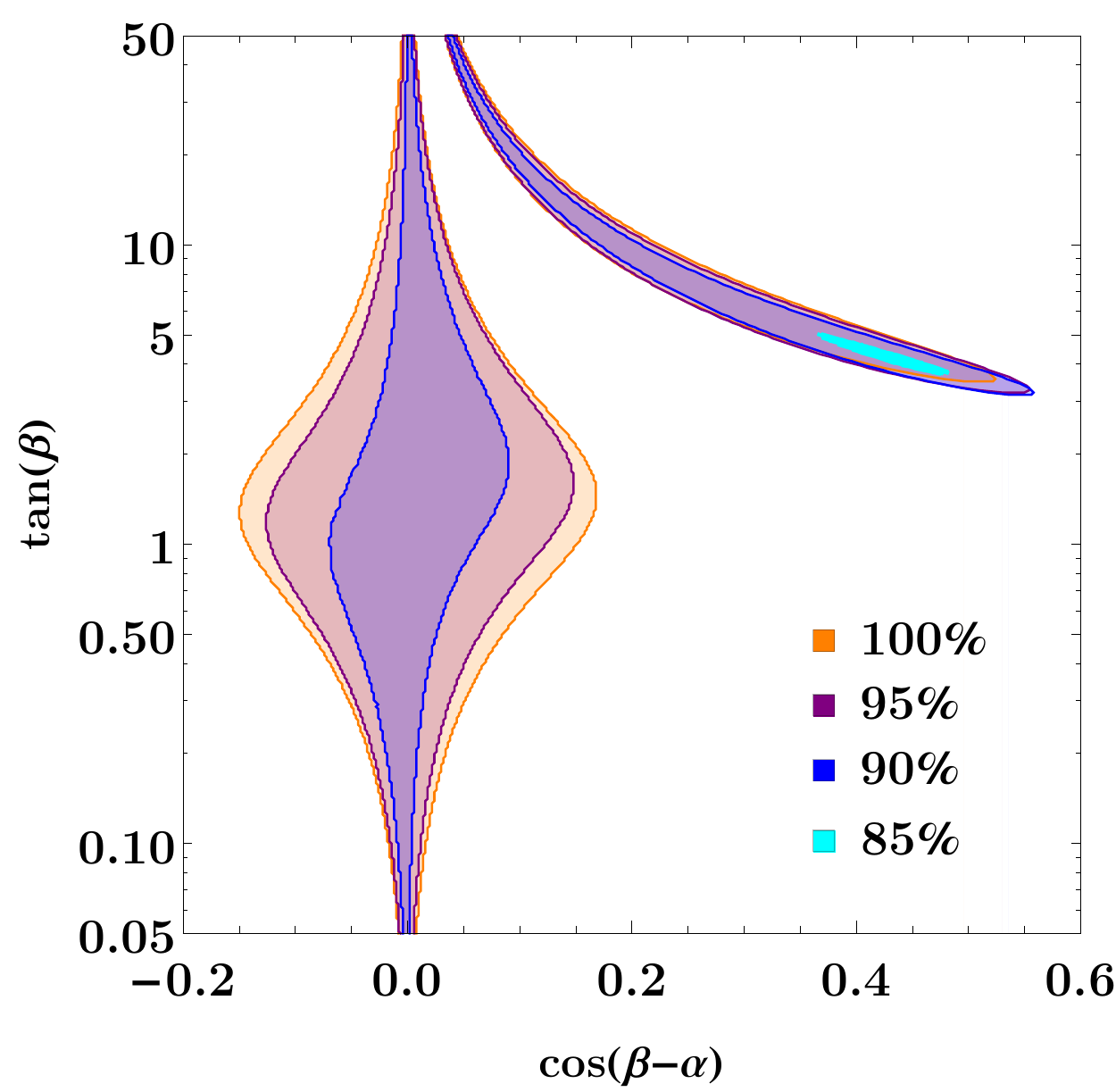}
\caption{Allowed regions in the $\tan \beta - \cos \left( \beta - \alpha \right)$ plane, in the model without $(q\tau)$-$(\mu e)$ mixing, for different values of $r \equiv \left( v_q^2 + v_\tau^2 \right)^{1/2} / v$, namely $r=1$ in orange, $r=0.95$ in purple, $r=0.90$ in blue and $r=0.85$ in cyan.}
\label{fig:plot1}
\end{figure}

We define the parameter $\mu_X$ as
\begin{equation}
    \mu_X \equiv \frac{\sigma(pp\rightarrow H) {\rm BR}(H \rightarrow X)} {\sigma(pp\rightarrow H)_{\rm SM} {\rm BR}(H \rightarrow X)_{\rm SM}}
\end{equation}   
and look at $X = gg, \mu\mu, \tau\tau, \bar{c}c, \bar{b}b, \bar{t}t,\gamma\gamma,\gamma Z, WW, ZZ$.
The results are in Figure~\ref{fig:plot1}, where we have plotted, in the usual way for 2HDMs, the allowed region in the $\tan\beta-\cos(\beta-\alpha)$ plane. We require all $\mu_X$ to be consistent with unity within 20\% at 95\% CL, which is a rough approximation to
the precision of current data.~\footnote{We are looking in the context of the lepton-specific 2HDM - but now the combination of vacuum values, $(v_q^2+v_\tau^2)^{1/2}$ no longer is equal to the Standard Model vacuum value, $v$.}

We see that if the ratio of $(v_q^2 + v_\tau^2)^{1/2}$ to $v$ is less than $0.85$, that the entire parameter space practically disappears.   Thus much of the vev is saturated by $v_q$ and $v_\tau$. Clearly the coupling here to the muon vanishes and thus in the full model, the muonic decay of the Standard Model Higgs, if confirmed, will be a strong constraint. 

The shrinking of the parameter-space in the $\cos({\beta-\alpha})<0$ allowed region occurs mainly due to the combination of $g_{HVV}^2$, measured from Higgs production, and $g_{Hll}^2$, measured from Higgs decay. The shrinking of the parameter-space in the $\cos({\beta-\alpha})>0$ allowed region mainly occurs due to $g_{HQQ}^2$, from Higgs production, now combined with both $g_{Hqq}^2$ and $g_{Hll}^2$.

In itself, this benchmark model is phenomenologically unacceptable.     Each $2\times 2$ submatrix will have a zero eigenvalue in the pseudoscalar and in the charged scalar sectors, leading to two zero eigenvalues in each sector.  Only one can be absorbed by the $W$ and $Z$ gauge bosons.    The additional massless scalars arise due to an additional accidental $SU(2)$ symmetry.    Thus, there must be some off-diagonal terms. We can include these terms but assume they are small and do a perturbative expansion.      

For simplicity, let us add a single off-diagonal term, $\lambda_5^{q\mu}$.    This will allow for nonzero masses for the lightest charged and pseudoscalar Higgs\footnote{One can decouple the masses of the charged and pseudoscalar Higgs by adding a $\lambda_4^{q\mu}$ term and can easily satisfy any BFB concerns with a $\lambda_3^{q\mu}$ term.}.   This term will modify the Yukawa couplings of the Standard Model 125 GeV Higgs.   For the couplings of the quarks, for example, the Yukawa coupling $g_Y\bar{q}q\Phi_q$ is $\sqrt{2}m_q/v_q$.  Writing $\Phi_q= V_{11}h_1 + V_{12}h_2 + ...$, where $h_1$ is the 125 GeV Higgs, one sees that the coupling is modified by a factor of $\frac{v}{v_q} V_{11}$.   One can perturbatively calculate the eigenvalues and eigenvectors of the mass matrix and we find that
\begin{equation}V_{11} = 1-\frac{1}{2}\epsilon_1^2\Bigg[ \left(\frac{c_{34}s_{12}}{m_{h_1}^2-m_{h_3}^2}\right)^2 + \left(\frac{c_{12}s_{34}}{m_{h_1}^2-m_{h_4}^2}\right)^2 \Bigg],
\end{equation}
where $\epsilon_1 = \lambda_5^{q\mu} v_qv_\mu$, $c_{ij} = \cos\alpha_{ij}$ ($s_{ij} = \sin\alpha_{ij}$) and the masses are the masses of the neutral scalars.   The relevant point here is that $V_{11}$ is reduced, which counters the effect of the smaller $v_q$.  In order for the lightest charged Higgs to have an acceptable  mass, there is a minimum value of $\lambda_5^{q\mu}$, but the masses of the neutral scalars can be large enough that the reduction (proportional to $(v_\mu/m_{h_3})^2$) is quite small.

\subsection{The Aligned Model}

The full 4HDM has a large number of parameters in the scalar potential: 10 quadratic terms and 22 quartic terms.    Not surprisingly, many of these parameters will have little effect on phenomenology.    As noted earlier, the fact that the $125$ GeV Higgs has decays consistent with the Standard Model implies that multi-doublet models must be near the alignment limit in which the Standard Model Higgs interactions are unaffected.   From Appendix A, we see that this will occur if $\alpha_{1j}= \beta_j$.  Parameters that might be of phenomenological relevance are then the $\beta_j$, $\alpha_{23,24,34}$, the three $\gamma$ parameters, the three $\delta$ parameters, the four scalar masses, the three charged masses and the three pseudoscalar masses, in addition to the SM Higgs vev. Instead of the potential’s couplings, we can choose to describe the model in terms of the previously mentioned parameters and six additional parameters, namely the remaining six $m_{ij}^2$, giving a total of 29 parameters\footnote{With the addition of the three $\alpha$ parameters which are defined through the alignment limit, we get 32 parameters, just like the scalar potential.}.    As we will see, many of these parameters will not be relevant for particular processes.   

One might wonder about relaxing the alignment limit assumption.   Since the Higgs properties match Standard Model expectations, one would expect deviations from the alignment limit to be small (of the order of $10 \%$ or less).   Since we are not including other similar size effects, such as radiative corrections to scalar masses, we don't anticipate any substantial effects on our plots.

Choosing values for the rotation angles and the squared masses, it is possible to define the scalar, pseudoscalar, and charged squared-mass matrices as $M_{s,p,c}^2 = R^{-1} D_{s,p,c} R$, considering the corresponding $R$ matrix for each case and $D$ as the diagonal matrix with the squared masses in its entries. The quartic parameters of the Lagrangian can be expressed in terms of elements of such matrices, the vevs and the $m_{ij}^2$ parameters as the following:
\be
\begin{aligned}
\lambda_1^i=&\frac{1}{2 v_i^3} \left( v_i M_{s,ii}^2 + \sum_{j \neq i} v_j m_{ij}^2 \right)\, , \\ 
\lambda_3^{ij}=&\frac{1}{v_i v_j} \left(M_{s,ij}^2 - 2 M_{c,ij}^2 + m_{ij}^2 \right)\, , \\ 
\lambda_4^{ij}=&\frac{1}{v_i v_j} \left(2M_{c,ij}^2 - M_{p,ij}^2 - m_{ij}^2 \right)\, , \\ 
\lambda_5^{ij}=&\frac{1}{v_i v_j} \left(M_{p,ij}^2 - m_{ij}^2 \right)\, , \\ 
\end{aligned}
\ee
in which $i,j=q,e,\mu,\tau$. In the 2HDM limit, these equations give rise to the well-known expressions for the $\lambda$ parameters in terms of masses, angles, the electroweak vev $v$ and the soft-breaking terms $m_{ij}^2$ \cite{Branco:2011iw, Branchina:2018qlf}.
For every possible set of parameters, we require the following:
\begin{itemize}
    \item The bounded-from-below conditions are satisfied.
    \item The perturbativity condition that the absolute values of $\lambda$ parameters are less than $4\pi$ is maintained.
    \item The previous condition also applies to Yukawa couplings.
    \item The values of the S and T parameters are within the range given by Eq.~(\ref{eq:oblique-values}).
    \item Charged Higgs masses must exceed $80$ GeV~\cite{ALEPH:2013htx}.
    \item Contributions from the charged scalars to the loop-induced Higgs diphoton decay $h \rightarrow \gamma \gamma$ are compatible with experimental bounds. This is achieved by checking the value of the diphoton signal strength $\mu_{\gamma \gamma}$~\cite{Bhattacharyya:2014oka,Djouadi:2005gj} for each set of parameters.
    \item Bounds coming from new physics contributions to $B$ meson oscillations, $\Delta{M_{B_{d,s}}}$, as well as $K$ mesons, $\Delta{M_K}$, are within the experimental allowed range for each case~\cite{Workman:2022ynf, FlavourLatticeAveragingGroupFLAG:2021npn}. Such nonstandard contributions come from charged scalars through one-loop processes~\cite{Chakraborti:2021bpy, Enomoto:2015wbn}.
    
    \item Contributions to $b\rightarrow s\gamma$~\cite{Enomoto:2015wbn}, again from charged Higgs particles, are acceptable.  In the Type II 2HDM, this gives the strongest constraint on charged Higgs bosons.
    \item  At the LHC, CMS~\cite{CMS:2018rmh} has searched for a heavy neutral Higgs decaying into $\tau$ pairs.    Although done in the context of the MSSM, the results are very similar in this model (with adjusted Yukawa couplings, of course) and the production cross-section times branching ratio varies from 10 pb to 10 fb over the range of masses from 150 GeV to 1000 GeV.  More recently, ATLAS~\cite{ATLAS:2020zms} has done a similar analysis.   Note that one usually assumes that the decay into top quarks will dominate for masses above 350 GeV, but that might not be the case here due to the lepton-specific nature of the model.   We impose these experimental bounds on our parameter-space, which, up to small differences due to form factors, apply to neutral scalars and pseudoscalars.
    \item  Finally, we can consider LHC direct searches for heavy charged Higgs bosons.    Searches fall into two categories - those in which the charged Higgs mass is greater than $m_t + m_b$ and those in which it is less. 
 \begin{itemize}
     \item If it is greater, then the predominant decay mode will be into $t\bar{b}$, except for the narrow window of parameter-space in which the charged Higgs in question has essentially zero overlap with $\Phi_q$.   The production cross-section for a charged Higgs mass of $200,300,600$ GeV is~\cite{Berger:2003sm} within a factor of 2 (scaling the Yukawa coupling appropriately to a lepton-specific or Type I model) of $0.4, 0.1, 0.01$ picobarns.  ATLAS~\cite{ATLAS:2021upq} has found bounds from Run II on the product of the production rate and the $H^+\rightarrow t\bar{b}$ branching ratio.   Their result is below our production cross-section by a factor of a few, and thus the model is not yet constrained by the non-observation at the LHC.  
     \item If the charged Higgs is lighter, then a major decay mode is into $\tau\nu_\tau$.   In this case the predominant production mode is through $t\rightarrow bH^+$.  Since top production is well understood, searches at ATLAS \cite{ATLAS:2018gfm} and CMS \cite{CMS:2019bfg} place bounds on ${\rm BR}(t\rightarrow bH^+) {\rm BR}(H^+\rightarrow \tau\nu_\tau)$.  This bound may not be too restrictive, since a charged Higgs that is either quarkphobic or leptophobic will not contribute and thus it will depend on mixing angles.   Nonetheless, we have incorporated the results of these searches in bounding our parameter-space.
 \end{itemize}
\end{itemize}

We will primarily focus on the lightest neutral scalar (other than the $125$ GeV Higgs), the lightest pseudoscalar and the lightest charged scalar.   Results from these scalars will also apply to the heavier scalars by appropriate choice of mixing angles (with the exception of heavy scalar decays into lighter scalars, which we will not consider).   The lepton-specific 2HDM has one scalar coupling to quarks and another to leptons.   The primary difference between our model and the lepton-specific model is that different scalars couple to the muon and the electron (note that the muon-specific model \cite{Abe:2017jqo,Ferreira:2020ukv} has the same scalar coupling to the quarks and the $\tau$, which is more like an extension of the type I 2HDM).   As a result, we will focus on decays involving muons and electrons.    

We first consider the decay of the lightest neutral scalar (other than the $125$ GeV Higgs, which has Standard Model couplings in the alignment limit) into electrons, muons and taus.  Since the heavier masses aren't relevant in the analysis, the parameter-space is substantially reduced.  
We consider two mass regions, in which the scalar mass is below and above $350$ GeV, respectively.   In the latter case, decays to top quarks can be substantial, even if the mixing angles are small.

As noted above, given the masses, soft-breaking mass parameters and mixing angles, the quartic couplings are determined.   We scan the full parameter space and check each of the conditions above.   Typically, we find several million parameter sets that are acceptable.  The results are plotted in Figure 2.    Note that in the Standard Model the branching ratio of the dimuon decay of the Higgs is $2\times 10^{-4}$ and this level (and somewhat below) is certainly experimentally accessible.    One can see that for a scalar mass below $350$ GeV, the dielectron decay branching ratio can be much, much larger than the Standard Model and the dimuon decay branching ratio can approach unity.  Above $350$ GeV, the opening of the top decay channel, even if the mixing angle is very small, substantially reduces the leptonic branching ratios.

\begin{figure}
\includegraphics[width=13.50cm]{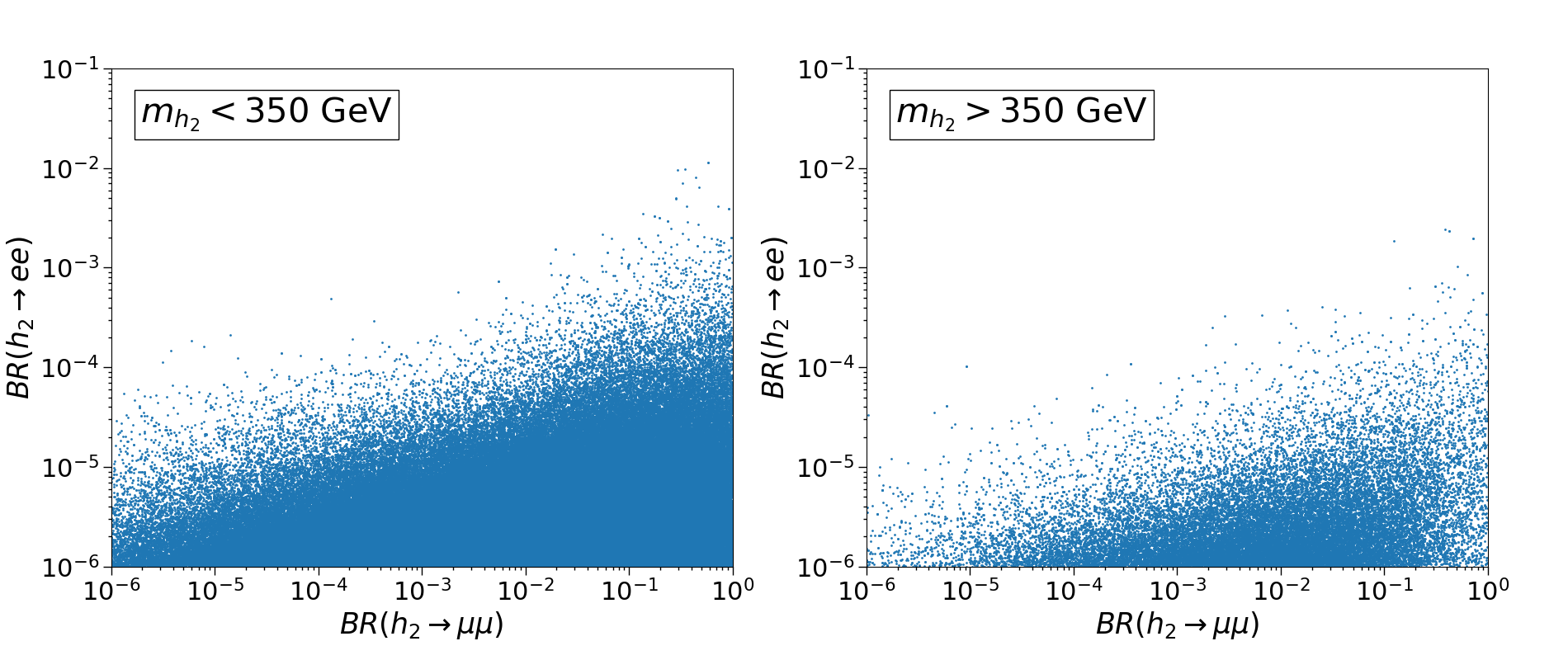}
\includegraphics[width=13.50cm]{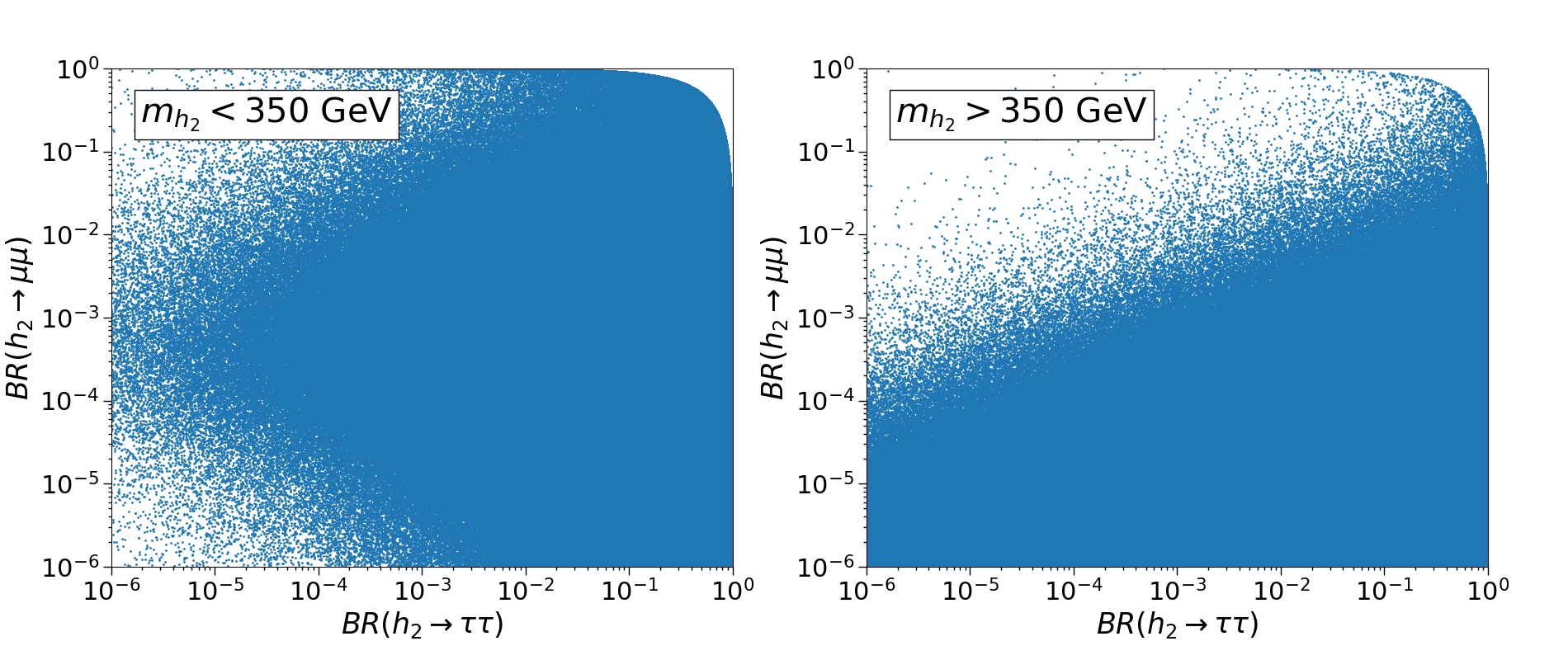}
    \caption{These scatterplots show allowed points for $h_2$ decays.   Results are shown for $h_2$ masses below $350$ GeV and above that mass scale (at which point the $\bar{t}t$ channel opens up).   The upper figures plot $ee$ and $\mu\mu$ decays and the lower figures plot $\mu\mu$ and $\tau\tau$ decays. The decay branching ratio of the SM Higgs to $\mu\mu$ is approximately $2\times 10^{-4}$.}
\label{fig:2}
\end{figure}

It is not surprising that this can occur.   If one chose parameters such that there was no mixing at all between $\Phi_{ee}$ and the other scalars, then the only decay of the $\Phi_{ee}$ would be into electrons. This would require extreme fine-tuning, since no symmetry will eliminate mixing in the quartic sector of the potential and even very small values of the quartic mixing terms would allow for other decays that could dominate.   Nonetheless, we see many sets of parameters for which the dielectron and dimuon decays of this lightest neutral scalar (other than the Standard Model Higgs) can be substantial.

In Figure 2, we also show the branching ratios to muons and  to taus.  Again, one can see that the absolute branching ratio to dimuons can be substantially more than that into two taus.    Thus, we find that searches for heavy neutral Higgs bosons decaying into leptons, which generally focus on tauonic decays, should also study muonic and electronic decays.

\begin{figure}
\includegraphics[width=13.50cm]{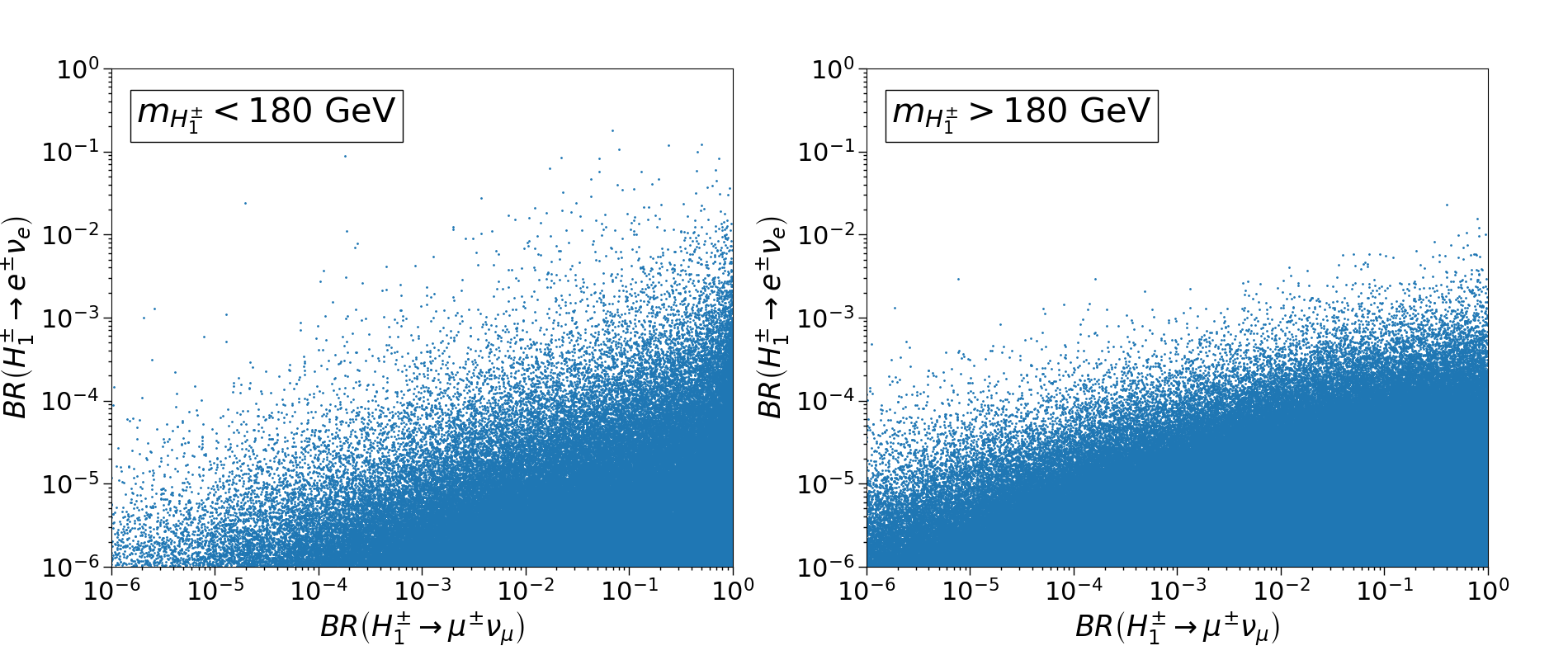}
\includegraphics[width=13.50cm]{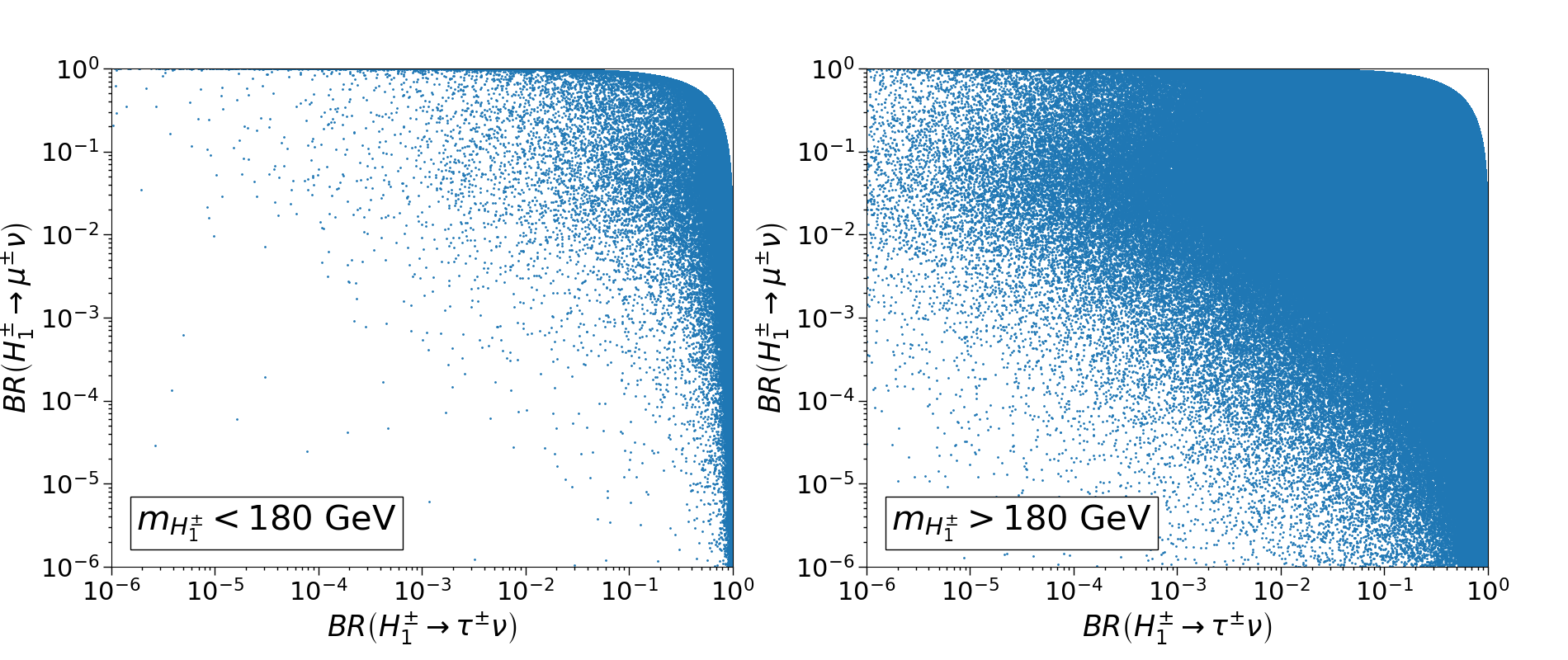}
\caption{These scatterplots show allowed points for $H^\pm$ decays.   Results are shown for $H^\pm$ masses below $180$ GeV and above that mass scale (at which point the $\bar{t}b$ channel opens up).   The upper figures plot $e\nu$ and $\mu\nu$ decays and the lower figures plot $\mu\nu$ and $\tau\nu$ decays.}
\label{fig:3}
\end{figure}

Since we are in the alignment limit, there is no three-point coupling of these scalars to two gauge bosons.    They could be produced in a collider through $WW$ or $ZZ$ fusion to two $\Phi$s.   The signature would be two electron-positron or muon pairs each coming from a $\Phi$.  The electron-positron pair rate will be smaller, but more distinctive.   While four lepton events have been searched for \cite{ATLAS:2021kog}, we know of no analysis of this particular signature.   An approximate production cross-section can be obtained by comparison with the inert doublet model \cite{Yang:2021hcu} which has a similar production process.  Typical production cross-sections at the LHC are approximately $0.5$ fb.    With an integrated luminosity of $3\ {\rm ab}^{-1}$, this means that branching fractions of $O(10^{-3})$ or less will be difficult to detect until the next generation colliders.

We have also studied the decays of the pseudoscalar into leptons and find very similar results.   For the charged Higgs decays, we show the branching ratios to $e\nu$ versus $\mu\nu$ decays in Figure 3 as well as the branching ratios for $\mu\nu$ versus $\tau\nu$ decays . Here, we consider mass ranges below and above $180$ GeV, at which point the $t\bar{b}$ opens up.  Note that there are more points in the region above $180$ GeV since below that mass a much higher proportion of points are experimentally excluded.  There is a large number of points in which the electronic decays are substantial and the muonic decay branching ratios can approach unity.

In Appendix C we show several benchmark points.    These points satisfy all of the various constraints listed earlier in this section.   For point S1, one can see that the $h_2\rightarrow\mu\mu$ branching ratio is almost $47\%$ and the electronic branching ratio is over $0.25\%$.  Clearly, the signature would most likely be two muon pairs, each coming from a neutral scalar, most of the other decays being tau pairs or $\bar{b}b$, with an occasional electron-positron pair.  In benchmark point S2, the dimuon decay of the scalar is smaller than that of the electron.  Here, one would see the ditau decays dominate, but the electron-positron decays might be measurable.   

We also consider some benchmark points for the lightest charged Higgs, looking at the region in which the mass is below $180$ GeV so the top-bottom channel is not available.    For point C1, the decay into muons is slightly bigger than the decay into taus, and the electronic decay is $0.2\%$.    For C2, the muon decay is the smallest of three branching ratios and the electron decay is as high as $1.7\%$.  Again, this shows that decays into muons and electrons might be much, much higher than in traditional 2HDMs.

\section{Conclusion}

It is often believed that all fermions of a given charge must couple to the same Higgs multiplet in order to avoid tree-level flavor-changing neutral currents.   However this is only true in the quark sector and need not be true in the lepton sector.   The quark mass matrix cannot be diagonal without eliminating CKM mixing, however the lepton mass matrix can be diagonal, since PMNS mixing can come from the superheavy Majorana neutrino sector.     We have studied a 4HDM in which one scalar doublet couples to quarks and the other three couple to the electron, muon and tau families, respectively.

There are numerous constraints on such a model, including bounded from below constraints, perturbativity, $S$ and $T$ parameters, the diphoton decay of the Higgs, limits from meson-antimeson oscillations, radiative $b$ decays and various LHC constraints from heavy scalar searches.    Scanning the parameter space, we find numerous acceptable points in which the dielectron and dimuon decays of the lightest neutral scalar (other than the $125$ GeV Higgs) can be much, much larger than expected.  The results for the lightest pseudoscalar and charged scalar are also presented.

Generally, searches for heavier Higgs bosons focus (in the lepton sector) on decays into $\tau$s.   However, this model shows that decays into electrons and muons can be substantial (and certainly easier to detect).    An interesting signature at either a linear collider or a hadron collider arises from vector boson fusion into two such Higgs bosons, each of which decays into an electron or muon pair.  We know of no bounds on such a process and hope to see searches in the near future.

\section*{Acknowledgments}

The work of MS and MK was supported by the National Science Foundation under Grant PHY-1819575.  The work 
of BLG is supported by Fundação para a Ciência e a Tecnologia (FCT, Portugal) through the PhD grant SFRH/BD/139165/2018 and the projects UIDB/00777/2020, UIDP/00777/2020, UIDB/00618/2020, UIDP/00618/2020, CERN/FIS-PAR/0019/2021 and CERN/FIS-PAR/0025/2021. BLG thanks the Fulbright Commission in Portugal and William \& Mary for support. MK thanks Pitt-PACC at the University of Pittsburgh for their hospitality. We thank Igor Ivanov for clarifying the symmetry group of the model, Arnab Dasgupta for coding help and suggestions and for useful discussions, and Pedro Ferreira for a helpful discussion of the lepton-specific 2HDM.
\pagebreak
\appendix
\section{Gauge Couplings}\label{label:appendixA}



\begin{center}
\begin{tabular}{| c || c |}

\hline

\multicolumn{2}{|c|}{\textbf{Trilinear Gauge Couplings $ZZh_i$ and $W^{\pm}W^{\mp}h_i$}} \\

\hline

 \parbox[t]{0.04\linewidth}{
 \centering
 $C_1$
 } & 
 \parbox[c][0.09\textheight]{0.93\linewidth}{
 \centering
 $
\text{c}_{12} \text{c}_{13} \text{c}_{14} \text{c}_{2} \text{c}_{3} \text{c}_{4} + \text{c}_{13} \text{c}_{14} \text{c}_{3} \text{c}_{4} \text{s}_{12} \text{s}_{2} + 
 \text{c}_{14} \text{c}_{4} \text{s}_{13} \text{s}_{3} + 
 \text{s}_{14} \text{s}_{4}
 $
 } \\

$C_2$ & 
 \parbox[c][0.1\textheight]{0.90\linewidth}{
 \centering
 $
 -\text{c}_{12} \text{c}_2 \text{c}_3 \text{c}_4 (\text{c}_{24} \text{s}_{13} \text{s}_{23} + 
    \text{c}_{13} \text{s}_{14} \text{s}_{24}) - 
 \text{c}_{23} \text{c}_{24} \text{c}_3 \text{c}_4 \text{s}_{12-2} - 
 \text{c}_{24} \text{c}_3 \text{c}_4 \text{s}_{12} \text{s}_{13} \text{s}_{23} \text{s}_2 - 
 \text{c}_{13} \text{c}_3 \text{c}_4 \text{s}_{12} \text{s}_{14} \text{s}_{24} \text{s}_2 + 
 \text{c}_{13} \text{c}_{24} \text{c}_4 \text{s}_{23} \text{s}_3 - 
 \text{c}_4 \text{s}_{13} \text{s}_{14} \text{s}_{24} \text{s}_3 + 
 \text{c}_{14} \text{s}_{24} \text{s}_4
 $
 } \\

 $C_3$ & 
 \parbox[c][0.225\textheight]{0.90\linewidth}{
 \centering
$
 -\text{c}_{12} \text{c}_{3} \text{c}_{4} [\text{c}_{13} \text{c}_{24} \text{c}_{2} \text{s}_{14} \text{s}_{34} + \text{s}_{23} (-\text{c}_{2} \text{s}_{13} \text{s}_{24} \text{s}_{34} + \text{c}_{34} \text{s}_{2}) + \text{c}_{23} (\text{c}_{34} \text{c}_{2} \text{s}_{13} + \text{s}_{24} \text{s}_{34} \text{s}_{2})] + \text{c}_{34} \text{c}_{4} [\text{c}_{2} \text{c}_{3} \text{s}_{12} \text{s}_{23} + \text{c}_{23} (-\text{c}_{3} \text{s}_{12} \text{s}_{13} \text{s}_{2} + \text{c}_{13} \text{s}_{3})] + \text{s}_{34} [\text{c}_{23} \text{c}_{2} \text{c}_{3} \text{c}_{4} \text{s}_{12} \text{s}_{24} + \text{c}_{3} \text{c}_{4} \text{s}_{12} \text{s}_{13} \text{s}_{23} \text{s}_{24} \text{s}_{2} - \text{c}_{24} \text{c}_{4} \text{s}_{13} \text{s}_{14} \text{s}_{3} - \text{c}_{13} \text{c}_{4} (\text{c}_{24} \text{c}_{3} \text{s}_{12} \text{s}_{14} \text{s}_{2} + \text{s}_{23} \text{s}_{24} \text{s}_{3}) + \text{c}_{14} \text{c}_{24} \text{s}_{4}]
$
 } \\

 $C_4$ & 
 \parbox[c][0.225\textheight]{0.90\linewidth}{
 \centering
$ 
-\text{c}_{2} \text{c}_{3} \text{c}_{4} \text{s}_{12} \text{s}_{23} \text{s}_{34} - \text{c}_{13} \text{c}_{24} \text{c}_{34} \text{c}_{3} \text{c}_{4} \text{s}_{12} \text{s}_{14} \text{s}_{2} + \text{c}_{34} \text{c}_{3} \text{c}_{4} \text{s}_{12} \text{s}_{13} \text{s}_{23} \text{s}_{24} \text{s}_{2} + \text{c}_{12} \text{c}_{3} \text{c}_{4} [-\text{c}_{13} \text{c}_{24} \text{c}_{34} \text{c}_{2} \text{s}_{14} + \text{c}_{34} \text{s}_{24} (\text{c}_{2} \text{s}_{13} \text{s}_{23} - \text{c}_{23} \text{s}_{2}) + \text{s}_{34} (\text{c}_{23} \text{c}_{2} \text{s}_{13} + \text{s}_{23} \text{s}_{2})] - \text{c}_{24} \text{c}_{34} \text{c}_{4} \text{s}_{13} \text{s}_{14} \text{s}_{3} - \text{c}_{13} \text{c}_{34} \text{c}_{4} \text{s}_{23} \text{s}_{24} \text{s}_{3} + \text{c}_{23} \text{c}_{4} [\text{c}_{34} \text{c}_{2} \text{c}_{3} \text{s}_{12} \text{s}_{24} + \text{s}_{34} (\text{c}_{3} \text{s}_{12} \text{s}_{13} \text{s}_{2} - \text{c}_{13} \text{s}_{3})] + \text{c}_{14} \text{c}_{24} \text{c}_{34} \text{s}_{4}
$
 } \\

\hline

\end{tabular}
\captionof{table}{$C_i$-factors of the trilinear gauge couplings $ZZh_i$ and $W^{\pm}W^{\mp}h_i$ as defined in Eq.~(\ref{eq:gauge-couplings}) in the main text. Here $\text{c}_{ij} = \cos\alpha_{ij}$ ($\text{s}_{ij} = \sin\alpha_{ij}$) and $\text{c}_{i} = \cos\beta_{i}$ ($\text{s}_{i} = \sin\beta_{i}$). In this notation, $\text{s}_{ij-k}$ stands for $\sin (\alpha_{ij}-\beta_k)$.} 
\end{center}
\pagebreak
\section{General Yukawa Couplings}\label{label:appendixB}



\begin{center}
\begin{tabular}{| c || c |}

\hline

\multicolumn{2}{|c|}{\textbf{General Yukawa Neutral Scalar}} \\

\hline

 \parbox[t]{0.04\linewidth}{
 \centering
 $\xi_h^{\text{ud}}$
 } & 
 \parbox[c][0.04\textheight]{0.93\linewidth}{
 \centering
 $\text{c}_{12} \text{c}_{13} \text{c}_{14} \; / \; \text{c}_2 \text{c}_3 \text{c}_4$
 } \\

 $\xi_h^e$ & 
 \parbox[c][0.04\textheight]{0.93\linewidth}{
 \centering
 $\text{s}_{12} \text{c}_{13} \text{c}_{14}  \; / \; \text{s}_2 \text{c}_3 \text{c}_4$
 } \\

 $\xi_h^{\mu }$ & 
 \parbox[c][0.04\textheight]{0.93\linewidth}{
 \centering
 $\text{s}_{13} \text{c}_{14} \; / \; \text{s}_3 \text{c}_4 $
 } \\

 $\xi_h^{\tau }$ & 
 \parbox[c][0.04\textheight]{0.93\linewidth}{
 \centering
 $ \text{s}_{14} \; / \; \text{s}_4$
 } \\
 
 \hline
 
 $\xi_{h_2}^{\text{ud}}$ & 
 \parbox[c][0.04\textheight]{0.93\linewidth}{
 \centering
 $-\left(\text{c}_{23} \text{c}_{24} \text{s}_{12} + \text{c}_{12} \left(\text{c}_{24} \text{s}_{13} \text{s}_{23} + \text{c}_{13} \text{s}_{14} \text{s}_{24}\right)\right) \; / \; \text{c}_2 \text{c}_3 \text{c}_4$
 } \\

 $\xi_{h_2}^e$ & 
 \parbox[c][0.04\textheight]{0.93\linewidth}{
 \centering
 $ \left(\text{c}_{12} \text{c}_{23} \text{c}_{24} - \text{s}_{12} \left(\text{c}_{24} \text{s}_{13} \text{s}_{23} + \text{c}_{13} \text{s}_{14} \text{s}_{24}\right)\right) \; / \; \text{s}_2 \text{c}_3 \text{c}_4$
 } \\

 $\xi_{h_2}^{\mu }$ & 
 \parbox[c][0.04\textheight]{0.93\linewidth}{
 \centering
 $\left(\text{c}_{13} \text{c}_{24} \text{s}_{23} - \text{s}_{13} \text{s}_{14} \text{s}_{24}\right) \; / \; \text{s}_3 \text{c}_4 $
 } \\

 $\xi_{h_2}^{\tau}$ & 
 \parbox[c][0.04\textheight]{0.93\linewidth}{
 \centering
 $\text{c}_{14}  \text{s}_{24} \; / \; \text{s}_4$
 } \\
 
 \hline
 
 $\xi_{h_3}^{\text{ud}}$ & 
 \parbox[c][0.04\textheight]{0.93\linewidth}{
 \centering
 $\left(\text{s}_{12} \left(\text{c}_{34} \text{s}_{23} + \text{c}_{23} \text{s}_{24} \text{s}_{34}\right) \! - \! \text{c}_{12} \left(\text{c}_{13} \text{c}_{24} \text{s}_{14} \text{s}_{34} + \text{s}_{13} \left(\text{c}_{23} \text{c}_{34} \!- \! \text{s}_{23} \text{s}_{24} \text{s}_{34}\right)\right)\right) / \text{c}_2 \text{c}_3 \text{c}_4$
 } \\

 $\xi_{h_3}^e$ & 
 \parbox[c][0.04\textheight]{0.93\linewidth}{
 \centering
 $ - \! \left(\text{c}_{12}\left(\text{c}_{34} \text{s}_{23} \! + \! \text{c}_{23} \text{s}_{24} \text{s}_{34}\right) \! + \! \text{s}_{12} \left(\text{c}_{13} \text{c}_{24} \text{s}_{14} \text{s}_{34} \! + \! \text{s}_{13} \left(\text{c}_{23} \text{c}_{34} \! + \! \text{s}_{23} \text{s}_{24} \text{s}_{34}\right)\right)\right) \! / \text{s}_2 \text{c}_3 \text{c}_4$
 } \\

 $\xi_{h_3}^{\mu}$ & 
 \parbox[c][0.04\textheight]{0.93\linewidth}{
 \centering
 $ \left(-\text{c}_{24} \text{s}_{13} \text{s}_{14} \text{s}_{34} + \text{c}_{13} \left(\text{c}_{23} \text{c}_{34} - \text{s}_{23} \text{s}_{24} \text{s}_{34}\right)\right) \; / \; \text{s}_3 \text{c}_4$
 } \\

 $\xi_{h_3}^{\tau}$ & 
 \parbox[c][0.04\textheight]{0.93\linewidth}{
 \centering
 $\text{c}_{14} \text{c}_{24} \text{s}_{34} \; / \; \text{s}_4$
 } \\

\hline

 $\xi_{h_4}^{\text{ud}}$ & 
 \parbox[c][0.04\textheight]{0.93\linewidth}{
 \centering
 $\left(\text{s}_{12} \left(\text{c}_{23} \text{c}_{34} \text{s}_{24} - \text{s}_{23} \text{s}_{34}\right) \! - \! \text{c}_{12} \left(\text{c}_{13} \text{c}_{24} \text{c}_{34} \text{s}_{14} \! - \! \text{s}_{13} \left(\text{c}_{34} \text{s}_{23} \text{s}_{24} + \text{c}_{23} \text{s}_{34}\right)\right)\right) / \text{c}_2 \text{c}_3 \text{c}_4 $
 } \\

 $\xi_{h_4}^e$ & 
 \parbox[c][0.04\textheight]{0.93\linewidth}{
 \centering
 $ - \! \left(\text{c}_{12}\left(\text{c}_{23} \text{c}_{34} \text{s}_{24} \! - \! \text{s}_{23}\text{s}_{34}\right) \! + \! \text{s}_{12} \left(\text{c}_{13} \text{c}_{24} \text{c}_{34} \text{s}_{14} \! - \! \text{s}_{13} \left(\text{c}_{34} \text{s}_{23} \text{s}_{24} \! + \! \text{c}_{23} \text{s}_{34}\right)\right)\right) \! / \text{s}_2 \text{c}_3 \text{c}_4$
 } \\

 $\xi_{h_4}^{\mu}$ & 
 \parbox[c][0.04\textheight]{0.93\linewidth}{
 \centering
 $-\left(\text{c}_{24} \text{c}_{34} \text{s}_{13} \text{s}_{14} + \text{c}_{13} \left(\text{c}_{34} \text{s}_{23} \text{s}_{24} + \text{c}_{23} \text{s}_{34}\right)\right) \; / \; \text{s}_3 \text{c}_4$
 } \\

 $\xi_{h_4}^{\tau}$ & 
 \parbox[c][0.04\textheight]{0.93\linewidth}{
 \centering
 $\text{c}_{14} \text{c}_{24} \text{c}_{34} \; / \; \text{s}_4$
 } \\

\hline

\end{tabular}
\captionof{table}{General Yukawa couplings of the scalar Higgs particles to quarks and charged leptons, as defined in Eqs.~(\ref{eq:yukawas-scalar-pseudoscalar-def1}) and~(\ref{eq:yukawas-scalar-pseudoscalar-def2}) in the main text. Here $\text{c}_{ij} = \cos\alpha_{ij}$ ($\text{s}_{ij} = \sin\alpha_{ij}$) and $\text{c}_{i} = \cos\beta_{i}$ ($\text{s}_{i} = \sin\beta_{i}$).} 
\end{center}



\begin{center}
\begin{tabular}{| c || c |}

\hline

\multicolumn{2}{|c|}{\textbf{General Yukawa Pseudoscalar}} \\

\hline

\parbox[t]{0.07\linewidth}{
 \centering
 $\xi_{A_1}^{\text{q}}$
 } & 
 \parbox[c][0.04\textheight]{0.9\linewidth}{
 \centering
 $-\left(\text{c}_{23} \text{c}_{24} \text{s}_2 + \text{c}_2 \left(\text{c}_{24} \text{s}_3 \text{s}_{23}+\text{c}_3 \text{s}_4 \text{s}_{24}\right)\right) \; / \; \text{c}_2 \text{c}_3 \text{c}_4$
 } \\

 $\xi_{A_1}^e$ & 
 \parbox[c][0.04\textheight]{0.9\linewidth}{
 \centering
 $\left(\text{c}_2 \text{c}_{23} \text{c}_{24} - \text{s}_2 \left(\text{c}_{24} \text{s}_3 \text{s}_{23} + \text{c}_3 \text{s}_4 \text{s}_{24}\right)\right) \; / \; \text{s}_2 \text{c}_3 \text{c}_4$
 } \\

 $\xi_{A_1}^{\mu}$ & 
 \parbox[c][0.04\textheight]{0.9\linewidth}{
 \centering
 $\left(\text{c}_3 \text{c}_{24} \text{s}_{23}-\text{s}_3 \text{s}_4 \text{s}_{24}\right) \; / \; \text{s}_3 \text{c}_4$
 } \\

 $\xi_{A_1}^{\tau}$ & 
 \parbox[c][0.04\textheight]{0.9\linewidth}{
 \centering
 $\text{s}_{24} \text{c}_4 \; / \; \text{s}_4$
 } \\
 
 \hline
 
 $\xi_{A_2}^{\text{q}}$ & 
 \parbox[c][0.04\textheight]{0.9\linewidth}{
 \centering
 $\left(\text{s}_2 \left(\text{c}_{34} \text{s}_{23}+\text{c}_{23} \text{s}_{24} \text{s}_{34}\right) - \text{c}_2 \left(\text{c}_3 \text{c}_{24} \text{s}_4 \text{s}_{34} + \text{s}_3 \left(\text{c}_{23} \text{c}_{34}-\text{s}_{23} \text{s}_{24} \text{s}_{34}\right)\right)\right) \! / \text{c}_2 \text{c}_3 \text{c}_4$
 } \\

 $\xi_{A_2}^e$ & 
 \parbox[c][0.04\textheight]{0.9\linewidth}{
 \centering
 $-\!\left(\text{c}_2 \left(\text{c}_{34} \text{s}_{23}+\text{c}_{23} \text{s}_{24} \text{s}_{34}\right) \! + \! \text{s}_2 \left(\text{c}_3 \text{c}_{24} \text{s}_4 \text{s}_{34}+\text{s}_3 \left(\text{c}_{23} \text{c}_{34} \! - \! \text{s}_{23} \text{s}_{24} \text{s}_{34}\right)\right)\right) \! / \text{s}_2 \text{c}_3 \text{c}_4$
 } \\

 $\xi_{A_2}^{\mu}$ & 
 \parbox[c][0.04\textheight]{0.9\linewidth}{
 \centering
 $\left(-\text{c}_{24} \text{s}_3 \text{s}_4 \text{s}_{34}+\text{c}_3 \left(\text{c}_{23} \text{c}_{34}-\text{s}_{23} \text{s}_{24} \text{s}_{34}\right)\right) \; / \; \text{s}_3 \text{c}_4$ 
 } \\

 $\xi_{A_2}^{\tau}$ & 
 \parbox[c][0.04\textheight]{0.9\linewidth}{
 \centering
 $\text{c}_{24} \text{s}_{34} \text{c}_4 \; / \; \text{s}_4$
 } \\
 
 \hline
 
 $\xi_{A_3}^{\text{q}}$ & 
 \parbox[c][0.04\textheight]{0.9\linewidth}{
 \centering
 $\left(\text{s}_2 \left(\text{c}_{23} \text{c}_{34} \text{s}_{24}-\text{s}_{23} \text{s}_{34}\right)-\text{c}_2 \left(\text{c}_3 \text{c}_{24} \text{c}_{34} \text{s}_4-\text{s}_3 \left(\text{c}_{34} \text{s}_{23} \text{s}_{24}+\text{c}_{23} \text{s}_{34}\right)\right)\right) \! / \text{c}_2 \text{c}_3 \text{c}_4$
 } \\

 $\xi_{A_3}^e$ & 
 \parbox[c][0.04\textheight]{0.9\linewidth}{
 \centering
 $-\!\left(\text{c}_2 \left(\text{c}_{23} \text{c}_{34} \text{s}_{24} - \text{s}_{23} \text{s}_{34}\right) \! + \! \text{s}_2 \left(\text{c}_3 \text{c}_{24} \text{c}_{34} \text{s}_4 \! - \! \text{s}_3 \left(\text{c}_{34} \text{s}_{23} \text{s}_{24} + \text{c}_{23} \text{s}_{34}\right)\right)\right) \! / \text{s}_2 \text{c}_3 \text{c}_4$
 } \\

 $\xi_{A_3}^{\mu}$ & \parbox[c][0.04\textheight]{0.9\linewidth}{
 \centering
 $-\left(\text{c}_{24} \text{c}_{34} \text{s}_3 \text{s}_4 + \text{c}_3 \left(\text{c}_{34} \text{s}_{23} \text{s}_{24}+\text{c}_{23} \text{s}_{34}\right)\right) \; / \; \text{s}_3 \text{c}_4$
 } \\

 $\xi_{A_3}^{\tau}$ & 
 \parbox[c][0.04\textheight]{0.9\linewidth}{
 \centering
 $\text{c}_{24} \text{c}_{34} \text{c}_4 \; / \; \text{s}_4$
 } \\
 
\hline

\end{tabular}
\captionof{table}{General Yukawa couplings of the pseudoscalar Higgs particles to quarks and charged leptons, as defined in Eqs.~(\ref{eq:yukawas-scalar-pseudoscalar-def1}) and~(\ref{eq:yukawas-scalar-pseudoscalar-def2}) in the main text. Here $\text{c}_{ij} = \cos\gamma_{ij}$ ($\text{s}_{ij} = \sin\gamma_{ij}$) and $\text{c}_{i} = \cos\beta_{i}$ ($\text{s}_{i} = \sin\beta_{i}$).} 
\end{center}


\begin{center}
\begin{tabular}{| c || c |}

\hline

\multicolumn{2}{|c|}{\textbf{General Yukawa Charged}} \\

\hline

\parbox[t]{0.07\linewidth}{
 \centering
 $\xi _{H_1^+}^{\text{qLR}}$
 } & 
 \parbox[c][0.04\textheight]{0.9\linewidth}{
 \centering
 $-\left(\text{c}_{23} \text{c}_{24} \text{s}_2 + \text{c}_2 \left(\text{c}_{24} \text{s}_3 \text{s}_{23}+\text{c}_3 \text{s}_4 \text{s}_{24}\right)\right) \; / \; \text{c}_2 \text{c}_3 \text{c}_4$
 } \\

 $\xi _{H_1^+}^{\text{eL}}$ & 
 \parbox[c][0.04\textheight]{0.9\linewidth}{
 \centering
 $\left(\text{c}_2 \text{c}_{23} \text{c}_{24} - \text{s}_2 \left(\text{c}_{24} \text{s}_3 \text{s}_{23} + \text{c}_3 \text{s}_4 \text{s}_{24}\right)\right) \; / \; \text{s}_2 \text{c}_3 \text{c}_4$
 } \\

 $\xi _{H_1^+}^{\text{$\mu $L}}$ &
 \parbox[c][0.04\textheight]{0.9\linewidth}{
 \centering
 $\left(\text{c}_3 \text{c}_{24} \text{s}_{23}-\text{s}_3 \text{s}_4 \text{s}_{24}\right) \; / \; \text{s}_3 \text{c}_4 $ 
 } \\

 $\xi _{H_1^+}^{\text{$\tau $L}}$ & 
 \parbox[c][0.04\textheight]{0.9\linewidth}{
 \centering
 $\text{s}_{24} \text{c}_4 \; / \; \text{s}_4$
 } \\
 
 \hline
 
 $\xi _{H_2^+}^{\text{qLR}}$ & 
 \parbox[c][0.04\textheight]{0.9\linewidth}{
 \centering
 $\left(\text{s}_2 \left(\text{c}_{34} \text{s}_{23} + \text{c}_{23} \text{s}_{24} \text{s}_{34}\right) - \text{c}_2 \left(\text{c}_3 \text{c}_{24} \text{s}_4 \text{s}_{34} + \text{s}_3 \left(\text{c}_{23} \text{c}_{34}-\text{s}_{23} \text{s}_{24} \text{s}_{34}\right)\right)\right) \! / \text{c}_2 \text{c}_3 \text{c}_4$
 } \\

 $\xi _{H_2^+}^{\text{eL}}$ & 
 \parbox[c][0.04\textheight]{0.9\linewidth}{
 \centering
 $-\!\left(\text{c}_2 \! \left(\text{c}_{34} \text{s}_{23} + \text{c}_{23} \text{s}_{24} \text{s}_{34}\right) \! + \! \text{s}_2 \left(\text{c}_3 \text{c}_{24} \text{s}_4 \text{s}_{34} + \text{s}_3 \left(\text{c}_{23} \text{c}_{34}-\text{s}_{23} \text{s}_{24} \text{s}_{34}\right)\right)\right) \! / \text{s}_2 \text{c}_3 \text{c}_4$
 } \\

 $\xi _{H_2^+}^{\text{$\mu $L}}$ & 
 \parbox[c][0.04\textheight]{0.9\linewidth}{
 \centering
 $\left(-\text{c}_{24} \text{s}_3 \text{s}_4 \text{s}_{34}+\text{c}_3 \left(\text{c}_{23} \text{c}_{34}-\text{s}_{23} \text{s}_{24} \text{s}_{34}\right)\right) \; / \; \text{s}_3 \text{c}_4 $
 } \\

 $\xi _{H_2^+}^{\text{$\tau $L}}$ & 
 \parbox[c][0.04\textheight]{0.9\linewidth}{
 \centering
 $\text{c}_{24} \text{s}_{34} \text{c}_4 \; / \; \text{s}_4$
 } \\
 
 \hline
 
 $\xi _{H_3^+}^{\text{qLR}}$ & 
 \parbox[c][0.04\textheight]{0.9\linewidth}{
 \centering
 $\left(\text{s}_2 \left(\text{c}_{23} \text{c}_{34} \text{s}_{24} - \text{s}_{23} \text{s}_{34}\right) - \text{c}_2 \left(\text{c}_3 \text{c}_{24} \text{c}_{34} \text{s}_4 - \text{s}_3 \left(\text{c}_{34} \text{s}_{23} \text{s}_{24} + \text{c}_{23} \text{s}_{34}\right)\right)\right) \! / \text{c}_2 \text{c}_3 \text{c}_4$
 } \\

 $\xi _{H_3^+}^{\text{eL}}$ &
 \parbox[c][0.04\textheight]{0.9\linewidth}{
 \centering
 $\left(\text{c}_2 \left(\text{s}_{23} \text{s}_{34}-\text{c}_{23} \text{c}_{34} \text{s}_{24}\right) - \text{s}_2 \left(\text{c}_3 \text{c}_{24} \text{c}_{34} \text{s}_4 - \text{s}_3 \left(\text{c}_{34} \text{s}_{23} \text{s}_{24} + \text{c}_{23} \text{s}_{34}\right)\right)\right) \! / \text{s}_2 \text{c}_3 \text{c}_4$
 } \\

 $\xi _{H_3^+}^{\text{$\mu $L}}$ & 
 \parbox[c][0.04\textheight]{0.9\linewidth}{
 \centering
 $-\left(\text{c}_{24} \text{c}_{34} \text{s}_3 \text{s}_4 + \text{c}_3 \left(\text{c}_{34} \text{s}_{23} \text{s}_{24}+\text{c}_{23} \text{s}_{34}\right)\right) \; / \; \text{s}_3 \text{c}_4 $
 } \\

 $\xi _{H_3^+}^{\text{$\tau $L}}$ & \parbox[c][0.04\textheight]{0.9\linewidth}{
 \centering
 $\text{c}_{24} \text{c}_{34} \text{c}_4 \; / \; \text{s}_4$
 } \\
 
\hline

\end{tabular}
\captionof{table}{General Yukawa couplings of the charged Higgs particles to quarks and leptons, as defined in Eqs.~(\ref{eq:yukawas-charged-def1}) and~(\ref{eq:yukawas-charged-def2}) in the main text. Here $\text{c}_{ij} = \cos\delta_{ij}$ ($\text{s}_{ij} = \sin\delta_{ij}$) and $\text{c}_{i} = \cos\beta_{i}$ ($\text{s}_{i} = \sin\beta_{i}$).} 
\end{center}
\pagebreak
\section{Benchmark Points}\label{label:appendixC}

\begin{center}
\begin{tabular}{| c || c || c |}

\hline

\parbox[t]{0.33\linewidth}{
 \centering
Scalar benchmark points
} & 
 \parbox[c][0.04\textheight]{0.33\linewidth}{
 \centering
 S1
 }& 
 \parbox[c][0.04\textheight]{0.33\linewidth}{
 \centering
 S2
 }

\\

\hline

\parbox[t]{0.33\linewidth}{
 \centering
$\beta_2 / \pi , \, \, \beta_3 / \pi , \, \, \beta_4 / \pi$
} & 
 \parbox[c][0.04\textheight]{0.33\linewidth}{
 \centering
 $0.05 , \, \, 0.16, \, \, 0.18$
 }& 
 \parbox[c][0.04\textheight]{0.33\linewidth}{
 \centering
 $0.04 , \, \,  0.14, \, \, 0.21 $
 }  

\\

\hline

\parbox[t]{0.33\linewidth}{
 \centering
$\alpha_{23} / \pi , \, \, \alpha_{24} / \pi , \, \, \alpha_{34} / \pi$
} & 
 \parbox[c][0.04\textheight]{0.33\linewidth}{
 \centering
 $-0.09, \, \, -1.00, \, \, -0.70 $
 }& 
 \parbox[c][0.04\textheight]{0.33\linewidth}{
 \centering
  $-0.02, \, \, -0.05, \, \, 0.10 $
 } \\

\hline

\parbox[t]{0.33\linewidth}{
 \centering
$\gamma_{23} / \pi ,\, \, \gamma_{24} / \pi ,\, \, \gamma_{34} / \pi$
} & 
 \parbox[c][0.04\textheight]{0.33\linewidth}{
 \centering
 $0.50 , \, \, 0.59, \, \,0.80 $
 }& 
 \parbox[c][0.04\textheight]{0.33\linewidth}{
 \centering
 $0.16, \, \, 0.52 , \, \,  0.39 $
 } \\

\hline

\parbox[t]{0.33\linewidth}{
 \centering
$\delta_{23} / \pi , \, \, \delta_{24} / \pi , \, \, \delta_{34} / \pi$
} & 
 \parbox[c][0.04\textheight]{0.33\linewidth}{
 \centering
 $0.08, \, \,  -0.26, \, \,-0.96 $
 }& 
 \parbox[c][0.04\textheight]{0.33\linewidth}{
 \centering
 $ 0.62, \, \,  -0.93, \, \,-0.95 $
  
 } \\

\hline

\parbox[t]{0.33\linewidth}{
 \centering
$m_{h_2} , \, \, m_{h_3} , \, \, m_{h_4} \, (\text{GeV})$
} & 
 \parbox[c][0.04\textheight]{0.33\linewidth}{
 \centering
 $269, \, \,  396, \, \, 483 $
 }& 
 \parbox[c][0.04\textheight]{0.33\linewidth}{
 \centering
   $175, \, \,  359, \, \, 360 $
 } \\

\hline

\parbox[t]{0.33\linewidth}{
 \centering
$m_{A_1} , \, \, m_{A_2} , \, \, m_{A_3} \, (\text{GeV})$
} & 
 \parbox[c][0.04\textheight]{0.32\linewidth}{
 \centering
 $439, \, \, 454, \, \, 484 $
 }& 
 \parbox[c][0.04\textheight]{0.33\linewidth}{
 \centering
 $265, \, \, 351, \, \, 369  $
 } \\

\hline

\parbox[t]{0.33\linewidth}{
 \centering
$m_{H_1^{\pm}} , \, \, m_{H_2^{\pm}} , \, \, m_{H_3^{\pm}} \, (\text{GeV})$
} & 
 \parbox[c][0.04\textheight]{0.33\linewidth}{
 \centering
 $438, \, \, 441, \, \, 443$
 }& 
 \parbox[c][0.04\textheight]{0.33\linewidth}{
 \centering
 $289, \, \, 352, \, \, 370$
  } \\

\hline

\parbox[t]{0.33\linewidth}{
 \centering
$m^2_{qe} , \, \, m^2_{q\mu} , \, \, m^2_{q\tau} \, (\text{GeV}^2)$
} & 
 \parbox[c][0.04\textheight]{0.33\linewidth}{
 \centering
 $-17700, \, 71700, \, -340000$
 }& 
 \parbox[c][0.04\textheight]{0.33\linewidth}{
 \centering
  $16000, \, -34600, \, -168000 $
 } \\

\hline

\parbox[t]{0.33\linewidth}{
 \centering
$m^2_{e\mu} , \, \, m^2_{e\tau} , \, \, m^2_{\mu\tau} \, (\text{GeV}^2)$
} & 
 \parbox[c][0.04\textheight]{0.33\linewidth}{
 \centering
 $-18600, \, 20700, \, -53600 $
 }& 
 \parbox[c][0.04\textheight]{0.33\linewidth}{
 \centering
 $14000, \, -31200, \, -57400 $
 } \\

\hline

\parbox[t]{0.33\linewidth}{
 \centering
$BR(h_2 \rightarrow ee)$
} & 
 \parbox[c][0.04\textheight]{0.33\linewidth}{
 \centering
  $2.72 \times 10^{-3}$
 }& 
 \parbox[c][0.04\textheight]{0.33\linewidth}{
 \centering
 $1.63 \times 10^{-4}$
 } \\

\hline

\parbox[t]{0.33\linewidth}{
 \centering
$BR(h_2 \rightarrow \mu \mu)$
} & 
 \parbox[c][0.04\textheight]{0.33\linewidth}{
 \centering
  $4.68 \times 10^{-1}$
 }& 
 \parbox[c][0.04\textheight]{0.33\linewidth}{
 \centering
$7.85 \times 10^{-6}$
 } \\

 \hline

\parbox[t]{0.33\linewidth}{
 \centering
$BR(h_2 \rightarrow \tau \tau)$
} & 
 \parbox[c][0.04\textheight]{0.33\linewidth}{
 \centering
 $1.22 \times 10^{-1}$
 }& 
 \parbox[c][0.04\textheight]{0.33\linewidth}{
 \centering
$7.42 \times 10^{-1}$
 } \\

\hline

\end{tabular}
\captionof{table}{Benchmark points for the leptonic decays of the lightest neutral scalar (other than the Standard Model Higgs) from Figure 2, for a $h_2$-mass range below $350~\text{GeV}$.} 
\end{center}

\begin{center}
\begin{tabular}{| c || c || c |}

\hline

\parbox[t][0.07\textheight]{0.33\linewidth}{
 \centering
Charged benchmark points
} & 
 \parbox[c][0.04\textheight]{0.33\linewidth}{
 \centering
C1
 }& 
 \parbox[c][0.04\textheight]{0.33\linewidth}{
 \centering
C2
 }

\\

\hline

\parbox[t]{0.3\linewidth}{
 \centering
$\beta_2 / \pi , \, \, \beta_3 / \pi , \, \, \beta_4 / \pi$
} & 
 \parbox[c][0.04\textheight]{0.3\linewidth}{
 \centering
 $0.05 , \, \, 0.05 , \, \, 0.09$
 }& 
 \parbox[c][0.04\textheight]{0.3\linewidth}{
 \centering
  $0.10 , \, \, 0.16 , \, \, 0.11$
 }  

\\

\hline

\parbox[t]{0.3\linewidth}{
 \centering
$\alpha_{23} / \pi , \, \, \alpha_{24} / \pi , \, \,\alpha_{34} / \pi$
} & 
 \parbox[c][0.04\textheight]{0.3\linewidth}{
 \centering
  $0.09 , \, \, 0.54 , \, \, 0.34$
 }& 
 \parbox[c][0.04\textheight]{0.3\linewidth}{
 \centering
 $0.20 , \, \, 0.88 , \, \, 0.72$
 } \\

\hline

\parbox[t]{0.3\linewidth}{
 \centering
$\gamma_{23} / \pi , \, \, \gamma_{24} / \pi , \, \,\gamma_{34} / \pi$
} & 
 \parbox[c][0.04\textheight]{0.3\linewidth}{
 \centering
  $-0.04 , \, \, 0.66 , \, \, 0.60$
 }& 
 \parbox[c][0.04\textheight]{0.3\linewidth}{
 \centering
 $0.68 , \, \, 0.50 , \, \, -0.52$
 } \\

\hline

\parbox[t]{0.3\linewidth}{
 \centering
$\delta_{23} / \pi , \, \, \delta_{24} / \pi , \, \,\delta_{34} / \pi$
} & 
 \parbox[c][0.04\textheight]{0.3\linewidth}{
 \centering
  $-0.98 , \, \, 0.00 , \, \, -0.36$
 }& 
 \parbox[c][0.04\textheight]{0.3\linewidth}{
 \centering
 $1.00 , \, \, 0.00 , \, \, 0.77$
 } \\

\hline

\parbox[t]{0.3\linewidth}{
 \centering
$m_{h_2} , \, \, m_{h_3} , \, \, m_{h_4} \, (\text{GeV})$
} & 
 \parbox[c][0.04\textheight]{0.3\linewidth}{
 \centering
 $127 , \, \, 187 , \, \, 208 $
 }& 
 \parbox[c][0.04\textheight]{0.3\linewidth}{
 \centering
 $180 , \, \, 237 , \, \, 240 $
 } \\

\hline

\parbox[t]{0.3\linewidth}{
 \centering
$m_{A_1} , \, \, m_{A_2} , \, \, m_{A_3} \, (\text{GeV})$
} & 
 \parbox[c][0.04\textheight]{0.3\linewidth}{
 \centering
 $131 , \, \, 179 , \, \, 244 $
 }& 
 \parbox[c][0.04\textheight]{0.3\linewidth}{
 \centering
 $161 , \, \, 172 , \, \, 173 $
 } \\

\hline

\parbox[t]{0.3\linewidth}{
 \centering
$m_{H_1^{\pm}} , \, \, m_{H_2^{\pm}} , \, \, m_{H_3^{\pm}} \, (\text{GeV})$
} & 
 \parbox[c][0.04\textheight]{0.3\linewidth}{
 \centering
  $164 , \, \, 172 , \, \, 229 $
 }& 
 \parbox[c][0.04\textheight]{0.3\linewidth}{
 \centering
  $158 , \, \, 181 , \, \, 234 $
  } \\

\hline

\parbox[t]{0.3\linewidth}{
 \centering
$m^2_{qe} , \, \, m^2_{q\mu} , \, \, m^2_{q\tau} \, (\text{GeV}^2)$
} & 
 \parbox[c][0.04\textheight]{0.3\linewidth}{
 \centering
 $-14800 , \, \, -17400 , \, \, 6210 $
 }& 
 \parbox[c][0.04\textheight]{0.335\linewidth}{
 \centering
 $57000 , \, \, -127000 , \, \, -15100 $
 } \\

\hline

\parbox[t]{0.3\linewidth}{
 \centering
$m^2_{e\mu} , \, \, m^2_{e\tau} , \, \, m^2_{\mu\tau} \, (\text{GeV}^2)$
} & 
 \parbox[c][0.04\textheight]{0.3\linewidth}{
 \centering
 $5880 , \, \, 22100 , \, \, 9060 $
 }& 
 \parbox[c][0.04\textheight]{0.3\linewidth}{
 \centering
  $-75600 , \, \, -9570 , \, \, 81300 $
 } \\

\hline

\parbox[t]{0.3\linewidth}{
 \centering
$BR(H_1^{\pm} \rightarrow e^{\pm}\nu_e)$
} & 
 \parbox[c][0.04\textheight]{0.33\linewidth}{
 \centering
 $2.24 \times 10^{-3}$
 }& 
 \parbox[c][0.04\textheight]{0.33\linewidth}{
 \centering
 $1.68 \times 10^{-2}$
 } \\

\hline

\parbox[t]{0.3\linewidth}{
 \centering
$BR(H_1^{\pm} \rightarrow \mu^{\pm}\nu_\mu)$
} & 
 \parbox[c][0.04\textheight]{0.3\linewidth}{
 \centering
 $5.36 \times 10^{-1}$
 }& 
 \parbox[c][0.04\textheight]{0.3\linewidth}{
 \centering
 $6.91 \times 10^{-3}$
 } \\

 \hline

\parbox[t]{0.3\linewidth}{
 \centering
$BR(H_1^{\pm} \rightarrow \tau^{\pm}\nu_\tau)$
} & 
 \parbox[c][0.04\textheight]{0.3\linewidth}{
 \centering
 $4.55 \times 10^{-1}$
 }& 
 \parbox[c][0.04\textheight]{0.3\linewidth}{
 \centering
 $5.23\times 10^{-1}$
 } \\

\hline

\end{tabular}
\captionof{table}{Benchmark points for the leptonic decays of the lightest charged scalar from Figure 3, for a $H_1^{\pm}$-mass range below $180~\text{GeV}$.} 
\end{center}

\pagebreak

\end{document}